\documentstyle[aaspp4,11pt]{article}
\begin{document}
\newcommand{\Deltas}{{\mbox{${\Delta S}$}}}
\newcommand{\vminusvhb}{{\mbox{${(V-V_{HB})}$}}}
\newcommand{\wca}{{\mbox{${W(Ca)}$}}}
\newcommand{\etal}{{\mbox{et al.}}}

\title{ The Abundance Spread Among Giants and Subgiants in
the Globular Cluster $\omega$ Centauri}

\author{Nicholas B.\ Suntzeff}
\affil{Dominion Astrophysical Observatory\\
Herzberg Institute of Astrophysics, National Research Council\\
5071 W. Saanich Road, Victoria, B.C. V8X 4M6, Canada\\
and\\
Cerro Tololo Inter-American Observatory\\
  National Optical Astronomy Observatories\altaffilmark{1}\\
  Casilla 603, La Serena, Chile\\ Electronic mail: nsuntzeff@noao.edu}
\author{Robert P.\ Kraft}
\affil{University of California Observatories/Lick Observatory\\
 Board of Studies in Astronomy and Astrophysics, University of California\\
 Santa Cruz, California 95064\\ Electronic mail: kraft@ucolick.org}
\altaffiltext{1}{ The National Optical Astronomy Observatories are 
  operated by the Association of Universities for Research in Astronomy, 
  Inc., under cooperative agreement with the National Science Foundation.}

\begin{abstract}

We present spectroscopic abundances and radial velocities for giant stars in
the Galactic globular cluster $\omega$ Centauri based on the \ion{Ca}{2}
infrared triplet.  Two samples of stars were observed: 234 stars at $M_V \sim
1.25$ on the lower giant branch at radial distances between 8\arcmin\ and
23\arcmin, and 145 stars at $M_V \sim -1.3$ at radial distances between
3\arcmin\ and 22\arcmin. We found 199 and 144 radial velocity members,
respectively, in the two samples. These samples were corrected for
evolutionary effects to provide an unbiased distribution of the underlying
stellar metallicity. We find $<v_r>=234.7 \pm 1.3, \sigma_{obs}=11.3$ km
s$^{-1}$ (bright sample), and $<v_r>=232.9 \pm 1.2, \sigma_{obs}=10.6$ km
s$^{-1}$ (faint sample). The statistical errors of the dispersions are less
than 1 km s$^{-1}$.

Previous metallicity studies found a non-gaussian metallicity distribution
containing a tail of metal-rich stars.  We confirm these results except our
unbiased cluster metallicity distributions are narrower.  They contain the
following key features: (1) No very metal-poor stars (2) a sudden rise in the
metal-poor distribution to a modal [Fe/H] value of --1.70 consistent with an
homogeneous metallicity unresolved at the 0.07 dex level, (3) a tail to higher
metallicities with more stars than predicted by simple chemical evolution
models, and (4) a weak correlation between metallicity and radius such that
the most metal-rich stars are concentrated to the cluster core. The unresolved
metal-weak tail implies that the gas out of which $\omega$ Cen formed was
well-mixed up to the modal metallicity of the cluster. Therefore, $\omega$ Cen
like other Galactic globular clusters, seems to have formed in a pre-enriched
and homogenized (up to the modal metallicity) environment.

The existence of a weak metallicity gradient supports the idea that $\omega$
Cen self-enriched, with the enriched gas sinking to the cluster center due to
gas dissipation processes. We also note, however, that the metal-rich stars
are more massive than the bulk of the stars in the cluster, and could also
have sunk to the center by dynamical mass segregation over the lifetime of the
cluster.

\end{abstract}


\section{Introduction}

$\omega$ Centauri is brightest globular cluster in the Milky Way Galaxy and
one of the most unusual.  It is the most massive cluster at $7 \times 10^6
M_{\sun}$ (Richer et al. 1991).  It has a very large spatial extent on the sky
with a core radius of 2.6\arcmin\ and a tidal radius of 45\arcmin ( Trager et
al. 1995).  It is also one of the dynamically {\it youngest} clusters with a
King concentration class of $c=1.24$ (Trager et al. 1995) and a half-mass
relaxation time of 5-10 Gyr (Binney \& Tremane 1987 with data from Webbink
1985).  Perhaps the most intensely studied peculiarity of $\omega$ Cen is the
substantial spread in Fe-peak metallicity, with a range of approximately 1
dex. In the early photometric studies, Cannon \& Stobie (1973) showed that the
color-magnitude (c-m) diagram of $\omega$ Cen had an unusually wide giant
branch, indicative of a heavy element abundance spread (cf. Norris \& Bessell
1975), and Butler \etal\ (1978) found an abundance range of [Fe/H] $\sim -0.6$
to --2.2 from a \Deltas\ -study of nearly half the cluster RR Lyraes. A
similar range of metallicity was found from the spread in the $(V-K)_0$
vs. $K_0$ diagram for 82 $\omega$ Cen giants (Persson \etal\ 1980).

High resolution spectroscopic studies of $\omega$ Cen giants, starting with
the pioneering work of Cohen (1981) and continuing to the present time (Mallia
\& Pagel 1981, Gratton 1982, Francois \etal\ 1987, Paltoglou \& Norris 1989,
Brown \& Wallerstein 1993) confirmed the existence of a wide spread in [Fe/H],
and in addition indicated that the detailed [element/Fe]-ratios are often not
those expected when one compares $\omega$ Cen giants with those found in
mono-metallic clusters. This in turn suggests that the nucleosynthetic history
of $\omega$ Cen is somehow different from that of other galactic globular
clusters. Details concerning these differences and what they imply about the
unique chemical evolution of $\omega$ Cen are the topics of several recent
studies (Norris \& Smith 1983, Paltoglou \& Norris 1989, Brown \& Wallerstein
1993), and especially a high resolution analysis of 40 $\omega$ Cen giants by
Norris \& Da Costa (1995).

Compared with these high resolution surveys, the present study has a more
limited objective, {\it viz.}, that of determining an unbiased measure of the
Fe-peak abundance spread in $\omega$ Cen. We attempt to derive the shape of
the [Fe/H]-distribution function using a complete sample of giants, drawn from
a well-chosen, but limited interval of absolute magnitude. We describe this
sample (actually two samples) in more detail in Section~\ref{secref1}. We note
here that previous investigators have chosen their stars so as to be
representative of the entire range of metallicities, and have not attempted to
construct a distribution function based on an unbiased sample. We hope to
recover here the shape of the distribution function, so that it may be
compared with simple evolutionary models such as the classical ``one-zone''
model of chemical evolution (e.g., Searle \& Sargent 1972, Hartwick 1976).

\section{Sample Selection; Abundance Method}
\label{secref1}

Our abundance procedure follows the technique developed by Olszewski \etal\
(1991), Armandroff \& Da Costa (1991), Armandroff \etal\ (1992), Da Costa
\etal\ (1992) and Suntzeff \etal\ (1992, 1993). We defer a discussion of
specific observational details to Section~\ref{secref2}, but describe the
principles of the method as follows. In spectra of $\omega$ Cen and certain
other mono-metallic cluster giants, we measure the pseudo-equivalent widths of
the two strongest \ion{Ca}{2} near-infrared triplet lines, $\lambda\lambda$
8542\AA\ and 8662\AA. We define a quantity $\wca = W(8542) + W(8662)$ which is
plotted as a function of \vminusvhb, where $V$ is the observed stellar visual
magnitude and $V_{HB}$ is the mean magnitude of the cluster horizontal branch
(HB).  The method depends on the well-known fact that in a $V$ vs $(B-V)$ c-m
diagram, giant branches become progressively taller (and bluer) with
decreasing [Fe/H]. For any mono-metallic cluster of a given [Fe/H], it turns
out that the curve relating \wca\ to \vminusvhb\ is a straight line with slope
that is independent of [Fe/H] and equal to 0.62 for stars lying above the HB,
ie., with \vminusvhb\ $< 0.00$ (see e.g., Figure 5 of Suntzeff \etal\ (1993);
hereafter S93). Interpolation of the $\omega$ Cen giants within the grid of
such straight lines for differing values of [Fe/H] leads directly to an
estimate of metallicity. The method has the obvious advantage that it is
independent of interstellar reddening and (often somewhat) imprecise
measurements of cluster distance moduli.

However, it is well known that the giant branch (GB) above the HB is bimodal:
most of these stars are ascending for the first time to the red giant tip
toward the He core flash, but about 20\% belong to the post-HB stage of
evolution, the so-called asymptotic giant branch (AGB) (Gingold 1974).
Generally these stars are brighter than their first giant branch counterparts
and the method described above will for these stars yield up [Fe/H]-values
that are somewhat too low. This presents a small problem in the analysis of
stellar systems lying at large distances (e.g., dwarf spheroidal satellites of
the Milky Way) in which only the brightest giants can be studied
spectroscopically. This is not necessarily a limitation, however, in the study
of the generally much nearer Galactic globular clusters.

Thus we have recalibrated the method using {\it subgiants}, i.e., cluster
stars lying {\it below} the level of the HB in the c-m array. These stars are
all on their first ascent up the GB.  That the method works empirically for
these subgiant stars will be demonstrated in Section~\ref{secref3}; suffice it
to say here that for subgiants, the slope of the lines relating W(Ca) to
\vminusvhb\ is different from that associated with the GB stars above the HB,
and thus the method had to be recalibrated from scratch. However, as a check
on the results based on subgiants, we also analyzed a sample of $\omega$ Cen
giants drawn from the tip of the GB.

The subgiant sample has a number of other advantages. The high density of
stars in a very limited magnitude and color range will reduce the effects of
systematic errors in converting line strength to the metallicity. Also, the
subgiant stars are hot enough to be free of TiO formation, which can severely
affect the measurement of metallicity in the \ion{Ca}{2} region (S93).

We selected $\omega$ Cen giant star candidates from the tables of $BV$
photometry published by Woolley (1966) (hereafter ROA). As noted above, two
groups of stars were selected: a ``subgiant sample'' (SG sample) with $14.8 <
V < 15.3$, and a ``bright giant sample'' (BG sample) with $12.2 < V <
12.8$. Different selection criteria were applied to the two samples.

The SG sample, with a mean $M_V \sim +1.25$, was drawn so as to avoid
selection effects that would bias the resulting metallicity distribution.  The
magnitude limits were set by two competing constraints. The fainter limit was
set both by the need to get a large sample in the small amount of observing
time allocated to the project and by the photometric accuracy of the source.
Cannon \& Stobie (1973) (see also Section~\ref{secref4}) found that the ROA
photometry is well-calibrated (up to a simple zero-point shift in the $B$ and
$V$ scales) with respect to their photoelectric photometry down to $V \sim
15.5$, but requires non-linear corrections below that level. The brighter
limit was set by the need to avoid the region on the GB where the stars slow
their evolution as they burn through the molecular weight discontinuity left
from main sequence hydrogen burning (Fusi Pecci \etal\ 1990). They found from
analysis of 13 cluster c-m diagrams, that the ``bump'' in the luminosity
function is brighter than the HB by about 0.45 mag at ${\rm [Fe/H]} = -2$ and
fainter than the HB by about 0.3 mag at ${\rm [Fe/H]} = -1$. With a mean HB
magnitude of $V \sim 14.5$, this sets the upper limit at about 14.8 in $V$.
We also made a cut in color at $(B-V) = 0.55$ to separate the GB from the
region of the HB. In $\omega$ Cen, there are almost no stars at this color, so
this provides a clean separation of the two evolutionary states.

To limit the number of field stars in the SG sample, we chose only stars from
the ROA study that were considered as zero-proper motion members (i.e., proper
motions of less than 100 in units given in the ROA work). We also excluded
stars which had no proper motions. In almost all cases, these stars were
within a cluster radius of 7.6\arcmin: this effectively defines our inner
radial limit for the sample. We excluded all stars with photometry listed as
uncertain in ROA (9\% of the sample). The stars with uncertain photometry had
the same radial distribution as the rest of the sample and their exclusion
should not bias the sample. Finally, we removed all stars that had any
companion (in the ROA lists) that was nearer than 5\arcsec\ of a given star;
such a companion would probably have contaminated the spectrum. A given star
of magnitude $V$ was excluded if there was any star within 5 arcsec that was
brighter than magnitude $V + 1.0$. This reduced the list by only 3\%. The
final list contained 250 stars, which is essentially an unbiased sample of
probable members between 8\arcmin\ and 23\arcmin\ from the cluster
center. Note that this is a radial region well outside the cluster half-light
radius of 4.8\arcmin\ (Trager et al. 1995).

The BG sample was created to tie into the previous abundance (both spectral
and photometric) studies, and to include a sample of stars that lie mostly at
smaller projected radii than the SG sample. We chose stars from the ROA lists
lying outside a radius of 3\arcmin, and which were subject to the following
restrictions: proper motion less than 127 (in ROA units) or {\it no} measured
proper motion; and stars with photometry not listed as ``uncertain''. As with
the SG sample, there was no red cutoff.  The final list contained 190 stars.
In addition to this sample, we also observed a list of $\sim 20$ bright
giants, observed as nightly standards by Dr.\ Pat Seitzer as part of his
program to measure the kinematics of the $\omega$ Cen cluster.  Observation of
these stars also permitted us to tie our abundance scale to that determined by
previous investigators, particularly the recent high resolution studies of
Paltoglou \& Norris (1989), Brown \& Wallerstein (1993) and Norris \& Da Costa
(1995). The Seitzer stars all have ROA numbers less than 233.

Both the SG and BG samples should be reasonably unbiased samples of the giants
in the corresponding radial annuli. They are not quite complete samples
however, because we have both excluded stars with poor photometry and adopted
a conservative cutoff in proper motion that may exclude some bona fide
members.

A c-m diagram showing the location of the two GB samples is shown in 
Figure~\ref{x1}.

To set up the abundance calibration of the \ion{Ca}{2} triplet lines, we
observed giants with luminosities comparable to those of our two $\omega$ Cen
samples in the globular clusters listed in Table~\ref{y1}.  The $V_{HB}$
values and reddenings were mostly taken from Armandroff (1989). The value of
$V_{HB}$ for NGC 3201 which was taken from Brewer et al. (1993).  The values
of $V_{HB}=14.54$ and $E(B-V)=0.11$ were taken from Butler et al (1978) which
are essentially identical to the independent work of Dickens \& Saunders
(1965).  We note here that the cluster [Fe/H] abundances are based on the Zinn
\& West (1984) scale. We discuss possible changes in this scale later.

\section{Observations and Data Reduction}
\label{secref2}
\subsection{Observational Techniques}

We observed these samples of stars with the CTIO 4m telescope and the Argus
multi-fiber system over a three-night period (24-26 Feb 1994 UT).  On the
first night, we observed for only 2.5 hours; the next two nights were
reasonably clear. Our observing and data reduction procedures are fully
described in S93, and we will not repeat them here. To briefly summarize, the
Argus system allows 24 objects to be observed simultaneously over a 50\arcmin\
region of the 4m telescope prime focus field. Twenty-four exposures of the sky
are also obtained with fibers set in close proximity to each stellar
object. The reduced spectra covered the wavelength range $\lambda\lambda
8200-8800$\AA\ with 1.2\AA\ pixel$^{-1}$ and 2.3 pixel resolution (R = 3100). A
few small differences in our reduction technique compared with that of S93
should be recorded.  First, because the stars in the present study are much
brighter than those dealt with by S93, we created a single median sky spectrum
per frame for each group of 24 object spectra.  Second, we applied only a
single set of nightly wavelength solutions to all 48 fibers, based on
exposures of a neon lamp which illuminated the dome white spot. The resulting
spectra are thus on a reasonably good relative wavelength scale, but could
suffer small shifts owing to slow nightly changes in the bench-mounted
spectrograph (these shifts are taken out, as described in the next
subsections).  Each program field was observed twice with the same exposure
times. All exposures of a given field were followed by exposures of a
comparison source taken at the same telescope position.  Multiple exposures
were co-added for measurement of the velocities and equivalent widths. The
individual exposures, however, were saved to allow one to make an estimate of
the errors.

The observed standard stars are listed in Table~\ref{y2}, and the observed
program stars are listed in Table~\ref{y3}a (BG sample) and \ref{y3}b (SG
sample). Table~\ref{y4} contains the observed stars in the BG and SG samples
that turned out to be non-members based on the radial velocities.  The
Tables~\ref{y2} and \ref{y3} contain the stars names, \vminusvhb\ magnitudes,
and dereddened $(B-V)_0$.  We list the average intensity in the co-added
spectra in units of ${\rm log}_{10}(e-)$ where ``e-'' refers to a detected
photo-electron.  Table~\ref{y3} also lists the radial distance (in arcminutes)
from the cluster center.  We observed 145 stars in the BG sample, 234 in the
SG sample, and 17 brighter giants for overlap with the Seitzer radial velocity
work. The number of observed stars is somewhat less than the full sample due
to fiber positioning limitations.

\subsection{Photometry of the sample}
\label{secref4}

In order to measure the stellar metallicities based on the \ion{Ca}{2} triplet
lines, we need $V$ photometry of only modest accuracy. An error of 0.1 mag in
$V$ will introduce an error of only 0.02 dex in [Fe/H] for a metal-poor star.
We have constructed average magnitudes from many sources in the literature.
For the BG sample, we have used the magnitudes listed in the ROA study, Eggen
(1972), Cannon \& Stobie (1973), and Lloyd Evans (1983a) as the primary
sources, and Bessell \& Norris (1976), Hesser, et al. (1977), Hawarden \&
Epps Bingham (1987), Martin (1981), and Alcaino \& Liller (1984) as secondary
sources.  There are well-known shifts between the various tables of
photometry.  We have calculated simple zero-point shifts of the primary
photometry lists to the system of Cannon \& Stobie (1973) as follows: ROA
(0.06,-0.07); Lloyd Evans 1983a (-0.03,0.00), Eggen 1972 (-0.06,0.03) where
the numbers refer to the additive constants applied to ($V$,$B-V$) in those
lists. There was not enough overlap among the secondary sources, so these were
used as given.  The averaged photometry from all the sources is listed in
Table~\ref{y3}a.  Many of the bright giants show large variations in $V$ as
noted in Table~\ref{y3}a.

We checked the ROA $V$ photometry for radial-dependent differences with the
photoelectric photometry. Other than the zero-points listed above, there was
no significant non-zero gradient from 5 to 21\arcmin\ for the 135 non-variable
red stars in common . However, the photometry of stars from 4 to 5 arcminutes
from the Lloyd Evans (1983a) work was $\sim 0.2$ mag {\em fainter} than the
ROA magnitudes.  Whether the ROA or the Lloyd Evans (1983a) photometry is in
error is difficult to say, but the $V$ magnitudes within 5 arcminutes of the
cluster center may be suspect.

To compare with the SG sample, there are 138 fainter stars with published
photoelectric (pe) photometry listed in Hawarden \& Epps Bingham (1987),
Cannon \& Stobie (1973), Martin (1981), and Norris \& Bessell (1975, 1977). If
we restrict the comparison to red stars with $B-V > 0.6$ and $13.1<V<15.3$,
the median differences are 0.04 mag or less in $V$ and 0.04 in $B-V$ between
the ROA and the other lists with dispersions (clipped at 3-sigma) of 0.07 mag
or less.  There was no non-zero gradient in the difference between the $V$
magnitudes from the ROA and pe photometry as a function of cluster radius from
10\arcmin\ to 23\arcmin.  We use the ROA photometry for the SG sample averaged
with the photometry from the other lists. The adopted photometry is listed in
Table~\ref{y3}b.

The photometry of the SG sample was checked with CCD data obtained on the CTIO
Curtis Schmidt telescope. A four frame mosaic of $\omega$ Cen in red and blue
colors was obtained with the Thomson 1024x1024 CCD (1.86 arcsec
pix$^{-1}$). The red data, taken with a 6840/95 filter, were analyzed with the
photometric packages DAOPHOT and ALLSTAR (Stetson 1994). The frames were star
subtracted, and then only the stars in the SG sample were added back in. These
frames were then measured with the DAOPHOT aperture photometry program with a
digital aperture of 7.5\arcsec\ radius. We calculated a simple linear
transformation between these red instrumental magnitudes and the corrected ROA
$V$ magnitudes for the SG sample. For 108 stars in overlap regions, the
average estimated (statistical) error in a single magnitude is 0.03 mag. The
transformation between red instrumental magnitude and $V$ had a dispersion of
0.07 mag, which is the same as the quoted accuracy of the ROA $V$ magnitudes
for single plate measurements.

The transformed red magnitudes (which we will call $V_R$) are an approximation
to the true $V$ magnitude, since [Fe/H] would be expected to affect $V$
differently than a redder color.  In Table~\ref{y3}b, we flag the four stars
where the ROA $V$ values differ by 0.2-0.3 mag from the $V_R$ CCD
photometry. No stars differed by more than 0.3 magnitude.  In the SG sample,
the two estimates of the $V$ magnitude agree to within 0.10 mag for 83\% of
the stars, and 94\% of the stars agree to within 0.15 magnitude.

There was a small gradient in $\Delta(V)$(CCD-ROA) of -0.012 mag arcmin$^{-1}$
from 8 to 17\arcmin\ radius and 0 -0.012 mag arcmin$^{-1}$ out to
24\arcmin. Evidently with the higher quality CCD magnitudes a small non-zero
gradient is evident in the ROA data, but the total effect of the gradient of
$\sim0.1$ mag will not affect the abundance results and we ignore it. 

We have also used the CCD frames to locate all the stars in the SG sample
which were observed.  The CCD xy positions were transformed to the ROA
astrometric positions with an accuracy of 0.4 arcseconds rms
(root-mean-square).  At the telescope, we noted a few cases where the there
was some ambiguity as to which star to assign to the fiber. We used the CCD
frames (after the run) to verify that the correct stars were indeed
observed. Only two stars were found to be incorrect. ROA 6124 was not found at
the telescope, yet it appears at the correct position in the CCD frames.
Presumably, the fiber positioner lost its absolute location for this star.
Star ROA 6176 was not located at the telescope nor on the CCD frame.

\subsection{ Measurements of Radial velocities and \protect{\ion{Ca}{2}} Triplet
Equivalent Widths}

The \ion{Ca}{2} triplet equivalent widths (W(Ca)) and radial velocities were
measured in the same manner as in S93, with a few differences noted below.
These values for the program and standard stars are listed in
Tables~\ref{y2},~\ref{y3}, and \ref{y4}.  The co-added spectra were used to
measure the values given in the tables, while the individual spectra were used
to calculate a standard deviation in the mean. This standard deviation, which
is listed as $\sigma$ in the tables, crudely represents the measurement error
in a single spectrum that was used in the co-addition to the final spectrum.
Also listed in the tables are the number of independent observations.

The standard clusters give the same W(Ca) as published in S93. We find the
following differences in the sense of (this work - S93): 47 Tuc, (0.12\AA,
0.16\AA, 11); M4, (0.07\AA, 0.12\AA, 18); and NGC 6397 (-0.03\AA, 0.09\AA,
20), where we list the mean difference, standard deviation , and number of
stars. The higher dispersion in the 47 Tuc stars is probably due to
variablility of stars on the RGB.

We reduced the radial velocities slightly differently from S93. A single
nightly wavelength calibration was used to correct all spectra to an
approximate linear wavelength scale. All stellar spectra were cross-correlated
with respect to three velocity templates (47 Tuc stars) exactly as in
S93. Fiber-to-fiber ``velocity'' differences were calculated by
cross-correlating the nightly twilight spectra with an averaged twilight
spectrum, and these offsets were subtracted from the stellar velocity in a
given fiber. At this point, all the velocities from spectra on a given frame
are on a relative velocity system. To bring all the frames onto a relative
velocity system, we calculated frame-to-frame offsets by cross-correlating the
sky spectrum (averaged over all 24 sky fibers) and subtracted this offset from
all the velocities in a given frame.  The same shifts were applied to the
comparison lamp arcs which were observed after the sequence of observations in
every field.  The final adjustment was made by comparing the observed
velocities with the following published velocities: 47 Tuc, Mayor et
al. (1983); M4, Peterson et al. (1995); and NGC 3201 C\^ote et al (1994).

On all three nights, a single field of bright giants (provided by Pat Seitzer)
was observed. The average velocity differences for the 24 stars were: night1 -
night2, (+0.5 1.2), and night1-night3 (+0.3,1.3) where we have listed the mean
difference and standard deviation of the mean in units of km s$^{-1}$. For a
field that was observed in 47 Tuc on nights 2 and 3, we find (-0.2, 1.9) km
s$^{-1}$.  Evidently the relative velocity zero point is accurate to better
than 1 km s$^{-1}$ for the three nights. The cross-correlation of the
comparison lamps yields an independent test of the relative velocity zero
points. The 24 spectra in the object fibers were co-added and
cross-correlated. The rms difference between the co-added comparison spectra
and a template was 0.5 km s$^{-1}$, consistent with a relative velocity zero
point accurate to better than 1 km s$^{-1}$.

In Table~\ref{y5}, we compare the velocities in Table~\ref{y2} with the
published values listed above. In addition, we compare the NGC 6397 velocities
with the values in S93. The comparison shows that for bright stars, the
individual velocities are accurate to 2 km s$^{-1}$ or better which we adopt as
the minimum error in the individual velocities.  The rms differences for the 4
cluster velocities given in Table~\ref{y5} imply an absolute velocity zero
point error of about 1 km s$^{-1}$.

Individual velocities for $\omega$ Cen stars have not be published, but mean
properties have been given by Meylan \& Mayor (1986) as summarized by Pryor \&
Meylan (1993) where they quote $<v_r>=232.2 \pm 0.7$ and $\sigma_{obs}=12.20
\pm 0.50$ km s$^{-1}$ for 318 stars in an annulus between 0.3\arcmin\ and
23.4\arcmin. For the obvious cluster members in our two samples, we find
$<v_r>=234.7 \pm 1.3, \sigma_{obs}=11.3$ km s$^{-1}, N=161$ for the BG+Seitzer
sample, and $<v_r>=232.9 \pm 1.2, \sigma_{obs}=10.6$ km s$^{-1}, N=199$ for
the SG sample. We have added the absolute velocity zero point error of 1 km
s$^{-1}$ to the statistical mean errors to derive the quoted errors in the
mean.

Excluding all stars with $\sigma(v_r) > 8$ km s$^{-1}$ in the SG sample does
not change these numbers significantly.  The statistical errors on the
dispersions ($\sigma_{obs} (2N)^{-0.5}$) are all less than 1 km s$^{-1}$.  The
slightly smaller velocity dispersions in our samples are probably due to the
falloff of dispersion with radius and the somewhat larger mean radius of our
samples.

The large radial velocity of $\omega$ Cen allows a clean separation between
members and non-members. All stars at greater than 3$\sigma_{obs}$ from the
cluster mean were considered non-members and are listed in Table~\ref{y4}. In
Figure~\ref{x6}, we show the c-m diagram for all the members and non-members
observed in $\omega$ Cen. Not surprisingly most non-members lie well off the
giant branch.

\subsection{Estimated Errors in W(Ca) and $v_r$}

The values for $\sigma$ given in Tables~\ref{y2} and \ref{y3} are estimators
for the errors in the derived quantities. The mean errors are formally
$\sigma/(N)^{1/2}$ but since N is typically 2, an individual mean error is not
well determined. A more robust way to estimate the errors is to correlate the
errors with the mean intensity and use the latter to estimate the error. In
Figure~\ref{x8} we plot all the $\sigma$ values for N=2 in our data set as a
function of the mean intensity in the co-added spectra. We also plot curves
which represent the median values of $\sigma$ from this and other Argus data
in our archives. These ensemble median values are given in Table~\ref{y7}. A
simple Monte Carlo calculation shows that for a normal distribution sampled
twice, $<\sigma>=0.79$ and ${\rm med}(\sigma)=0.67$ where $\sigma$ is the
dispersion calculated for two points. Therefore, the mean error of a
measurement based on the co-added spectrum should be ${\rm med}(\sigma) /
(0.67\times \sqrt{2}$) which is roughly the ${\rm med}(\sigma)$ given in
Table~\ref{y7}.  The mean intensity is not a perfect indicator of the errors
because the stars with truly large values of $\sigma$ in Tables~\ref{y2} and
\ref{y3} do have poorly determined velocities and equivalent widths.

As a compromise, we suggest the following estimator for the errors: the error
in a quantity is the larger of two values, ${\rm med}(\sigma)$ and
$\sigma/(2)^{1/2}$ for N=2. For the two samples in $\omega$ Cen, the median
errors in W(Ca) are 0.05 and 0.16\AA.

\section{Calibration and Results}
\label{secref3}
\subsection{The Reduced Equivalent Widths for the Two Samples}

Except in the case of 47 Tuc, two groups of stars were observed to mimic the
luminosity range of the $\omega$ Cen samples. The block of observing time
allocated to the project proved inadequate to permit the acquisition of a 47
Tuc sample analogous to the $\omega$ Cen SG sample; the effect of this on our
results should be slight, and is discussed a little later.

In Figure~\ref{x2}, we plot \wca\ as a function of \vminusvhb\ for the
calibrating clusters. The long straight solid lines are the best fit ridge
lines with constant slope of 0.62 transferred from Figure 5 of S93 for the
clusters M71, M4 and NGC 6397.  The dashed line is the fit to the NGC 3201
data from this paper with the same slope.  The solid and dashed lines, all
having a slope of 0.62 in this diagram, correspond to the stars of the BG
sample of each of the calibrating clusters. The dotted straight lines, on the
other hand, are ``by eye'' fits to the sample of SG stars of
Table~\ref{y2}. Note that the slopes of the dotted lines are essentially the
same (0.35), but differ from the slope associated with the bright star
samples. As noted above, we did not observe SG stars in 47 Tuc, but assumed by
analogy with the situation for the BG sample, that the 47 Tuc SG stars would
fall along a line of slope 0.35 in a position homeomorphic to the location of
the lines representing the other clusters.

For the BG sample, if we form the ``reduced'' equivalent widths of the two
strongest \ion{Ca}{2} lines as $W' = W(Ca) + 0.62(\vminusvhb)$, the parameter
$W'$ becomes an estimator of [Fe/H], as previously shown in S93, Armandroff
\etal\ (1992), and Armandroff \& Da Costa (1991).  The averaged values of $W'$
and the errors in the mean for the standard clusters based on the BG sample
are given in Table~\ref{y1}. The $<W'>$ values for 47 Tuc, M4, and NGC 6397 in
Table~\ref{y1} are essentially identical to the values for these clusters
listed in S93 of 4.75, 4.08, and 2.30\AA.

In analogy with the BG sample, we can also form a reduced equivalent width for
the SG sample, but with a different slope. If we force this reduced equivalent
width to be on the same scale as that of the brighter giants, we find $W' =
W(Ca) + 0.35(\vminusvhb) - 0.19 $. For the subgiant stars in M4, NGC 3201, and
NGC 6397, we find the following values of $<W'>$ for the SG sample: $4.11 \pm
0.02{\rm \AA}, 3.54 \pm 0.04{\rm \AA}, 2.36 \pm 0.03{\rm \AA}$, which are
essentially identical to the mean values in Table~\ref{y1}. We can treat the
$W'$ system for both the BG and SG samples as the same.

In Figure~\ref{x7} we plot the W(Ca) values for the $\omega$ Cen giants in the
two samples. The distribution of stars with respect to the ridge lines is very
similar in the two samples, and we can expect that both samples will give very
similar metallicity distributions.  There is a sharp cutoff at low metallicity
which is parallel to the ridge lines and most of the stars in the sample are
near this cutoff. There is also a less populated tail up to line strengths
significantly greater than that of 47 Tuc giants.  The fact that there is no
similar tail or even a single star at lower metallicities (in the SG sample)
strongly indicates that $\omega$ Cen did not form in situ with lower
metallicity gas, but instead, formed in the presence of mono-metallic gas as
did almost all the other galactic globular clusters. The four stars in the BG
sample that evidently have low metallicities may well be AGB stars, which lie
too bright with respect to RGB stars at the same $B-V$ colors.

\subsection{The Metallicity Calibration}

We have two available calibrations to convert $W'$ into [Fe/H]. The first
calibration is from S93 where we plot [Fe/H] as a function of $<W'>$ for
galactic globular clusters. We show this relationship in Figure~\ref{x3}. We
have drawn two line segments by eye to mimic the mean relationship. These
relationships are:

\noindent
${\rm [Fe/H]} = 0.37W' - 2.79$, $W' < 4.1${\rm \AA}\ and \\
${\rm [Fe/H]} = 0.73W' - 4.28$, $W' > 4.1${\rm \AA}.

The calibration on the metal-poor end is slightly different than the one given
in S93, because we have forced the calibration to be continuous at the break
point, rather than use least squares for the two segments which produces a
discontinuity at $W'=4.1$.

For the second calibration, we can use the actual [Fe/H] values of $\omega$
Cen giants from the high-dispersion work of Norris \& Da Costa (1995) for the
24 stars in common. In Figure~\ref{x4} we plot the relationship between
individual $W'$ values and the [Fe/H] and [Ca/H] for the stars in common. We
have plotted both [Fe/H] and [Ca/H] to illustrate the following point. The
$W'$ value for a star is a function of $T_{eff}$, log(g), broadening
parameters, [Ca/H], and the continuous opacity. By plotting along isochronal
sequences (as we do when we plot W(Ca) as a function of \vminusvhb) and taking
out the mean trend, we have formed what appears to be a functional dependence
of $W'$ on a single value - [Fe/H]. This is valid insofar as [Ca/Fe] is a
monotonic function of [Fe/H], and that [Ca/Fe] has the same relationship in
individual stars in $\omega$ Cen as the average trend seen in the globular
clusters. The Norris \& Da Costa (1995) work suggests that [Ca/Fe] is a
different function of [Fe/H] than the average trend for globular clusters in
that [Ca/Fe] remains high even for metal-rich stars. If this is the case, the
globular cluster calibration shown in Figure~\ref{x3} will be wrong for
metal-rich stars and it will {\it over-estimate} the true value of [Fe/H] by
up to 0.3-0.4 dex. In addition, the larger scatter seen in Figure~\ref{x4} for
the ($W'$,[Fe/H]) compared to ($W'$,[Ca/H]) may also indicate that there is
real scatter in the [Ca/Fe] relationship in $\omega$ Cen.

We have fit a functional form to the ($W'$,[Fe/H]) relationship shown in
Figure~\ref{x4} as

\begin{equation}
{\rm [Fe/H]} = -2.45{\rm \AA} + 0.223W' + 0.0152(W')^2, 2.6 < W' < 5.4{\rm
\AA}.
\label{eq1}
\end{equation}

The two values of [Fe/H] will be called ${\rm [Fe/H]}_{ZW}$ and ${\rm
[Fe/H]}_{NDC}$. One final refinement will be made to the derived
metallicities. In deriving [Fe/H] we have used the quantity \vminusvhb.  The
level of the HB however, depends on [Fe/H], so we must correct the quantity
\vminusvhb\ for [Fe/H], and rederive the metallicity. To do this, we have
adjusted $V_{HB}$ as

\begin{equation}
V_{HB} = 14.54 + 0.20({\rm [Fe/H]} + 1.7) 
\label{eq2}
\end{equation}

Here we have assigned the average value of $V_{HB}$ based on the RR Lyraes to
the metallicity of [Fe/H]=--1.7 (which is the modal value of [Fe/H] in
$\omega$ Cen), and assumed a slope of 0.20, which is a compromise value among
the slopes of the ($M_V$, [Fe/H]) relationship of RR Lyraes (Sandage \&
Cacciari 1990).  The values of $V_{HB}$ for $\omega$ Cen adjusted in this
manner have been used to recalculate \vminusvhb\ and the values listed in
Table~\ref{y3} have been corrected for this effect. The final values of [Fe/H]
listed in Table~\ref{y3} also reflect the adjustments in $V_{HB}$. The effect
of this refinement is small and affects only the metal-rich stars: the
adjustment makes the metal-rich star appear brighter with respect to $V_{HB}$,
and as can be seen in Figure~\ref{x2}, this will make the star more
metal-poor. The change is no larger than $\sim 0.1$ dex.

In Figure~\ref{x5} we illustrate the change in the appearance of the IR
\ion{Ca}{2} triplet lines in three stars drawn from the SG sample. The
equivalent widths of the triplet lines change by a factor of two in this
illustration, which corresponds to a change of a factor of 10 in [Fe/H].

We leave this section with a final comment on the metallicity scale. While we
have no independent spectroscopic evidence to decide which scale is correct,
the colors of the stars appear to support the [Fe/H]$_{NDC}$ scale. In
Figure~\ref{x12}, we plot the derived [Fe/H] values for the two scales as a
function of dereddened $(B-V)$. The $(B-V)_0$ for a star in this evolutionary
position should have essentially the same functional dependence on [Fe/H] as
$(B-V)_{0,g}$, which is the mean dereddened color of a globular cluster RGB at
the level of the horizontal branch. Figure~\ref{x12} shows that the
[Fe/H]$_{NDC}$ scale reproduces the mean trend of $(B-V)_{0,g}$ for globular
clusters better than the [Fe/H]$_{ZW}$ scale. The extremely high metallicities
in the [Fe/H]$_{ZW}$ scale are not confirmed by very red colors.

\section{The Metallicity Distribution Functions}

\subsection{Evolutionary Corrections to the Abundance Histogram}

Before we can discuss the metallicity distribution function, we need to make
one final correction to our samples.  By chosing samples of stars via cuts at
constant $V$, we will subtly bias our data due to the complex way in which
stars evolve as a function of metallicity. There are three major effects. 

The first effect is that stars of higher metallicity will have a larger
bolometric luminosity at a fixed $V$.  This is mainly due to the fact that the
bolometric correction (to $V$)) is only a weak function of [Fe/H] at a given
$(B-V)$ but is a strong function of (B-V) for the range in temperatures and
metallicities of the stars considered here. The metal-rich stars, which have
redder colors, have a larger bolometric correction and are intrinsically more
luminous at a given $V$.  In this case, the higher metallicity stars will
evolve more quickly through the cut in $V$ and therefore be underrepresented.

The second effect is for a given age of the underlying population, a higher
metallicity star will have a higher mass. For instance, at $M_V=1.25$ which is
representative of the SG sample, the mass increases from 0.82 to
$0.87M_{\sun}$ as the metallicity increases from [Fe/H]=--1.78 to --0.78 for
14 Gyr isochrones (Bergbusch \& VandenBerg 1992). Since mass functions,
defined as $ {\rm d}N = AM^{-x}{\rm d} \log(M)$ typically have positive values
of $x$ (Richer et al. 1991), the metal-rich stars will again be
underrepresented with respect to a fixed number of main sequence stars.

The third effect is due to mass segregation where more massive stars sink to
the center of the cluster. We will come back to this point later.  Other
effects, such as variable mass functions (as a function of [Fe/H]) are further
complications which will not be considered here.

The first two effects can be corrected for by numerical integration of
theoretical luminosity functions, as given in Bergbusch \& VandenBerg (1992).
The second effect however, requires that we define a base population to which
we will correct all the masses and this correction will require the adoption
of a mass function. We cannot use the subgiant branch or main sequence turnoff
as the base, since the number of stars per unit interval in these sequences
are also a function of metallicity. We will use the brightest unevolved main
sequence stars, with masses between 0.6 and 0.7$M_{\sun}$ as the base.  The
choice of the brightest unevolved main sequence stars minimizes the effects of
the adopted mass function.  We calculate the correction factor to the raw
histogram as the ratio of the number of stars between 0.6 and 0.7$M_{\sun}$ to
the number of stars in the given $V$ interval.  We will use an apparent
distance modulus to the cluster of 13.83. This was calculated using the
$V_{HB}$([Fe/H]) relationship given above with a modal cluster metallicity of
--1.7.

The choice of the mass function is not clear. Richer et al. (1991), in their
Figure 10, present mass functions for many clusters, including $\omega$ Cen.
The four metal-poor clusters in that figure have rather flat mass functions
from 0.7 to 0.4$M_{\sun}$. A simple fit to these 4 clusters yields an average
value of $x=0.8 \pm 0.3$. However, the measured mass function at 0.8$M_{\sun}$
in $\omega$ Cen is extremely steep ($x \sim 4$) until 0.6$M_{\sun}$ where it
flattens to $x=-1$. Dynamical processes can sink more massive stars to the
cluster center, but given the distance of these $\omega$ Cen fields from the
cluster center (5 and 9$r_c$) and the long half-mass relaxation time of the
cluster, this seems unlikely.  It should be noted that the mass function at
the high mass end is based on a very few number of stars, and we suspect that
the problem lies with counting statistics.  We will adopt a more typical mass
function with $x=1.0$.

We find that the correction factor is relatively independent for ${\rm [Fe/H]}
< -1.0$ but is much more dependent on the mass function at higher
metallicities. For instance, the correction factor at [Fe/H]=--0.65 rises from
15\% to 26\% as the mass function power law increases from $x=1$ to 2.  The
correction factors are plotted in Figure~\ref{x9}.

In Figure~\ref{x10} we present the metallicity histograms, corrected for the
evolutionary effects, and in Table~\ref{y6} we give the basic cluster
statistics for $\omega$ Cen. 

\subsection{Comparison of the SG and BG samples }

Both the cluster averages given in Table~\ref{y6} and the histograms for the
two samples shown in Figure~\ref{x10} show that the two samples have very
similar distributions.  From smoothed histogram fits, we estimate the modal
value as [Fe/H]=--1.70 for the two distributions.  There is a sharp rise from
lower metallicities to the modal value of the distribution, and a less steep
decline to higher metallicities. A tail with a few stars extends to very high
metallicities.

There are, however, small differences in the two sample distributions.  If we
pare the BG sample to only include stars at cluster radii greater than
8\arcmin\ to be consistent with the SG sample, we find that a two-sided
Kolmogorov-Smirnov (K-S) test (Nemec \& Harris 1987, Press, et al. 1992) gives
a probability of the rejection of the hypothesis that the two samples are
drawn from the same parent population at 80\%, independent of which
metallicity scale is chosen. The sense of the difference is that the BG sample
has a small number of metal-rich stars that are much more metal-rich than the
SG sample.

It is difficult to pinpoint the reason for the small differences in the
distributions. The calibrations in Figures~\ref{x3} and~\ref{x4} cover the
range in line strengths observed. One may suspect that because the conversion
of W(Ca) to W' shown in Figure~\ref{x2} does not extend to metallicities
higher than those in M4 for the SG sample, the metal-rich end of the SG
metallicity scale may be suspect. More calibration data will be needed to
resolve this problem.

One may also suspect the calibration of the BG sample. There is a trend of
increasing W(Ca) as a star ascends the RGB shown in Figure~\ref{x2}. But as
shown by Erdelyi-Mendes \& Barbuy (1991), for [Fe/H]=--1, there is almost {\it
no} gravity effect on W(Ca) at log(g)=0, while there is a trend of increasing
line strength with decreasing $T_{eff}$. The trend seen in Figure~\ref{x2} is
due to mostly the change in $T_{eff}$ on the RGB. According to Norris \& Da
Costa (1995), the metal-rich stars have unusual [${\alpha}$/Fe] ratios. As
discussed by VandenBerg (1991), different abundances of the intermediate
$\alpha$-elements will affect the position of the isochrones in lower
temperature stars.  Thus, the use of the correction of W(Ca) as a function of
\vminusvhb\ which is based on normal globular cluster iso-abundance sequences
may also be suspect. Isochrone models which decouple CNO, [${\alpha}$/Fe], and
[Fe/H] are needed to investigate this effect.

Since the difference between the two samples is really seen
only in the few metal-rich stars, we must also be concerned that the line
strengths, either in the high-dispersion data of Norris and Da Costa (1995) or
this work, could be affected by molecular formation or non-LTE ionization
effects. For instance, the $\alpha$-element abundances tabulated in the Norris
and Da Costa (1995) work rely on neutral lines which could be affected by
non-LTE ionization effects in the cooler stars or errors in the $T_{eff}$
calibration.  New high S/N high-dispersion spectra of the metal-rich giants
would be very useful in assessing these effects.
\subsection{Metallicity gradient in $\omega$ Cen}

In Figure~\ref{x11} we plot the radial distribution of the metallicity in
$\omega$ Cen. In a general sense, this diagram shows there is a large
dispersion in metallicities at all radii, and there is no large gradient in
the average metallicity as a function of cluster radius.

There are also some indications of possible trends buried within the large
metallicity dispersion. There appear to be a few very metal-poor giants in the
BG sample that are not seen in the SG sample. While these could be metal-poor
giants, it is more likely they are AGB stars, which will have artificially low
W(Ca) values due to their position in the c-m diagram (S93). We have flagged
these stars (ROA 342,383,429,435) in Table~\ref{y3}a. We remove these stars
from the following discussions.

There appears to be a trend in that the most metal-rich giants in the BG
sample lie a small radii. This trend is not seen in the SG sample, but the SG
sample does not probe the inner part of the cluster where most of these stars
lie.  If we cut the BG sample in two at a given [Fe/H], we can compare the
radial distributions with a K-S test. We find that if the cut is made in the
range of $-1.1 <$ [Fe/H] $< -1.3$, the two distributions are different at the
70\% level.  The effect disappears if cuts are taken at lower metallicities.
Thus we have evidence, albeit weak, that there is a some segregation in
metallicity as a function of radius.

According to Freeman (1985), giants lying beyond the limits of the ROA catalog
have systematically weaker CN bands than the majority of stars inside the $r_t
= 22$ arcmin radius. Weaker CN bands are believed to be an indication of
reduced metallicity, although the size of the effect has not been calibrated.
A comparison of CN band strengths in clusters such as M13 (at ${\rm [Fe/H]} =
-1.6$) with M92 (at ${\rm [Fe/H]} = -2.2$) suggests that over this 0.6 dex
decline in metallicity, CN bands essentially disappear (cf.  Suntzeff 1981,
Carbon \etal\ 1982). Thus over the metallicity range encountered among
$\omega$ Cen giants we might expect that a change in [Fe/H] as small as 0.3
dex could induce a significant change in the strength of the CN bands.
However, other effects related to evolutionary state may also be at work (see
review by Kraft 1994), and even among giants in the same evolutionary state
and metallicity, but in different clusters, CN band strengths can differ
(Langer \etal\ 1992).  The interpretation of this effect and its influence on
the abundance distribution derived here, although probably small, is not
entirely clear.

\subsection{Discussion}

The metallicity histograms shown in Figure~\ref{x10} verify the general shape
of the metallicity histograms given by Butler et al. (1978), Norris (1980),
Persson, et al. (1980), and Da Costa \& Villumsen (1981): a sharp rise at a
low metallicity and a tail of stars to higher metallicity, although our
unbiased sample shows that there are relatively few stars on the metal-rich
tail. Our data do not support the existence of a rather large population of
very metal-poor stars found by Butler, Dickens \& Epps (1978) in their
\Deltas\ study of 58 RR Lyraes in $\omega$ Cen, where they found that one
quarter of these stars had ${\rm [Fe/H]} \la -1.90$. The derived metallicities
in their study, which were based on image-tube spectra, had rms errors of 0.4
dex and presumably the low-metallicity stars are the low-metallicity tail of
the error distribution. A new study of the metallicities of RR Lyraes in
$\omega$ Cen is called for, since CCD spectra can give abundances based on
\Deltas\ values good to 0.15 dex in [Fe/H] (Suntzeff et al. 1994). 

Is there a metal-poor tail? If we fit the SG distribution with a gaussian from
the metal-poor tail up to the modal value of [Fe/H]=--1.70, we find a
dispersion of 0.07 dex in [Fe/H] for 69 stars. If we only take the 36 stars
with W(Ca) errors less than median value of $\sigma({\rm W(Ca)}$ of 0.16\AA,
we still find the same dispersion. Note that the relations above give
$\sigma({\rm[Fe/H]}) \approx 0.35 \sigma({\rm W(Ca)}$) at the metal-poor end.
The errors show that we are just at the limit of detecting a real metallicity
spread on the metal-poor tail. We can conclude that the metal-poor tail of
$\omega$ Cen stars is consistent with these stars having formed in a
proto-cluster gas clouds with a metallicity spread of less than about 0.07 dex
in [Fe/H].  This level of chemical homogeneity is typical for other globular
clusters (Suntzeff 1993). Some clusters have limits of inhomogeneity half this
value, but these clusters generally have had many fewer stars observed at much
higher signal-to-noise.  Better spectra (with multiple observations) could
improve the estimate of the metal-poor tail homogeneity in $\omega$ Cen. It
should be noted that we cannot use the BG sample for this because of the AGB
contamination.

The overall metallicity distribution can be compared to a model of chemical
evolution (Searle \& Sargent 1972) as conveniently parametrized by Hartwick
(1983) and Pagel (1992).  Pagel describes this ``Simple Model'' as a
``one-zone model treated in the instantaneous recycling approximation with
constant yields for 'primary' nucleosynthesis products from short-lived
massive stars..,'' of the functional form:

\begin{equation}
\frac{{\rm d} s}{{\rm d} \ln z} \propto \frac{z}{y} e^{-(z-z_0)/y)}
\end{equation}

Here $z$ is the instantaneous mass fraction of the metals, $s$ is the mass of
stars with abundances $\le z$, $y$ is the yield, and $z_0$ is the initial
abundance of the gas. The distribution is 0 for $z > \ln (m/g)$ where $g$ is
the mass of gas and dust, and $m=g+s$ is the total mass of the system.  We
have modified this distribution following Zinn (1978) and Bond (1981) to
include a non-zero starting metallicity. This distribution peaks at $z=y$
(provided that $z_0 < y$) which leads one to assume that the yield is very
roughly the solar metallicity for simple models of disk evolution, which is
verified from calculations of nucleosynthetic yields where $y \sim 0.01$
(Matteucci 1983).

The metallicity distribution for $\omega$ Cen stars peaks at a much lower
metallicity ($z =0.0003$) implying a much lower yield is needed. Low yields
are calculated for metal-poor stars ($y \sim 0.001$; Matteucci 1983) but this
is still not low enough.  The yield can also be lowered by assuming very steep
mass function on the main sequence or by introducing gas loss at a rate
proportional to the star formation rate. This latter refinement by Hartwick
(1978) is suggested by the simple idea that the number of number of supernovae
should be related directly to the star-formation rate.  In this case, the
distribution has the same functional form but the total mass $m$ is no longer
constant, and the yield is replaced by an effective yield which is a factor of
$(1 + \Lambda)$ smaller than the actual yield, where $\Lambda$ is the ratio of
mass loss to star formation (in Pagel's notation).

An example fit to the data using a Simple Model is shown in Figure~\ref{x13}.
Here we have used an initial metallicity of [Fe/H]=--1.78, and an effective
yield of 0.00012. We have convolved the model with the observed median error
of 0.07 dex, and binned the model data for comparison. Larger effective yields
will improve the fit to the metal-rich tail but will broaden the peak well
past the observed limits. Smaller effective yields will narrow the model peak
(and improve the fit in this Figure) but will produce even fewer metal-rich
stars.  

If the Simple Model is correct, then the poor fits as evidenced in
Figure~\ref{x13} can be explained in two ways. We can lower the yield ( thus
underproducing still further the metal-rich stars) and postulate a second
generation of stars formed with a higher mean metallicity to account for the
metal-rich tail. Pre-enriched gas falling back on the cluster after the
initial star formation wave is a natural explanation for this. We can also
increase the yield slightly to fit the metal-rich tail (which will broaden the
metallicity peak), and postulate a homogeneous initial population of stars at
[Fe/H] $\approx -1.7$. Both scenarios require two generations of stars.

If the low effective yields are interpreted as cluster mass loss, then
$\Lambda > 10$ implying almost the whole cluster mass is lost (Zinn
1978). Smith (1984) discusses the fate of a proto-cluster in the presense of
severe mass loss. If the mass loss is impulsive such that the cluster cannot
adjust adiabatically, it will disrupt with only one-half the cluster mass
driven off. The time-scale for impulsive processes must be less than the
adiabatic time scale, which is on order of a few crossing times or roughly 3
Myr for $\omega$ Cen. If the mass loss is slow enough that the cluster can
adjust adiabatically, then the cluster will bloat out such that the cluster
radius is inversely proportional to the mass.  While this may be the case for
a dwarf spheroidal galaxy (which was the goal of the modeling done by Smith),
this can hardly be the case for a globular cluster (Gunn 1980).

Smith does point out that if roughly half of the cluster formed stars {\it
before} chemical enrichment happened, then the remaining gas could self-enrich
and ultimately be swept out of the cluster by supernovae, without having the
cluster disrupt. If this gas fraction leaves slowly enough that the cluster
can adjust adiabatically, then the cluster will expand by a modest amount. In
support of this latter scenario, he shows that $\omega$ Cen has the lowest
central concentration for a cluster of its size.

However, at the point where we begin to look into detail about the time scales
for mass ejection and star formation, the utility of the Simple Model and its
assumption of instantaneous star formation must be questioned. The fact that
the free-fall time for a cluster (Fall \& Rees 1985), the crossing time, and
the typical age for the massive stars are all in the range of $10^6 - 10^7$yr
argues that the chemical enrichment must be tied to the dynamical evolution of
the young cluster.  One must look towards more dynamical models of cluster
formation, which will certainly have more free parameters. There are many
models of cluster formation which try to realistically predict the environment
of star formation in a proto-cluster (see, for instance, Fall \& Rees 1985,
Cayrel 1986, Morgan \& Lake 1989, Brown et al. 1991 and 1995, and Vietri \&
Pesce 1995). As an example, we note the model of Cayrel, as recast by Brown et
al. In this model, the proto-cloud center collapses quickly forming a very
large OB association. Supernovae and stellar winds from the massive stars
produce an annulus of well-mixed enriched material.  This annular region forms
a second generation of stars which can form a globular cluster if the
thermalized stellar velocities are small enough for the stars to be held in
the gravitational potential of the cluster.  While this model attempts to
explain the chemical homogeneity of globular clusters through the turbulent
mixing in the annular shell, any explanation for a metallicity spread in
$\omega$ Cen would be a complication on top of the theory.

We have also detected a weak metallicity gradient in $\omega$ Cen. This lends
support to an extended time scale for formation of stars, since the metal-rich
gas would have time to dissipate its kinetic energy prior to forming stars.
However, there is also another interpretation to the metallicity gradient.
Previously we noted that the metal-rich stars would be more massive than
coeval metal-poor stars. We would then expect the more massive metal-rich
stars to sink to the cluster center, on the time scale of the half-mass
relaxation time. The relaxation time for the cluster is very large, perhaps
half the cluster age, so the effectiveness of the gravitational sinking is
hard to estimate without detailed multi-mass King models. Richer et al (1991)
do give equilibrium multi-mass models showing a large amount of mass
segregation for the most massive stars. The existence of a metallicity
gradient outside the half-mass radius where the dynamical time-scales should
be much longer than the cluster age would provide strong evidence for
dissipative energy loss and an extended time scale for chemical enrichment.

In summary, we find that the functional form of the Simple Model of chemical
enrichment gives the same qualitative shape as the metallicity distribution in
$\omega$ Cen, but in detail, this model underestimates the number of stars on
the metal-rich tail.  The lack of a resolved metal-poor tail means that
$\omega$ Cen shares the property with the other Galactic globular clusters
that the bulk of the cluster stars formed in a chemically homogeneous
environment.  Any reasonable fit to the data with the Simple Model requires
extremely low yields which in turn implies a cluster mass loss that will bloat
the cluster or even more likely cause it to be unbound. The concurrence of
important time scales for cluster and stellar formation argue that the Simple
Model is inadequate for predicting chemical evolution, and a more dynamical
model is required. The metallicity gradient, while supporting the idea of gas
dissipation processes leading to a more concentrated later generation of
metal-rich stars, also could be an outcome of the sinking of the more massive
(but coeval) metal-rich stars over the lifetime of the cluster.

\acknowledgments

We thank Patrick Seitzer for sharing his list of bright $\omega$ Cen giant
abundance standards and Pat C\^ote for a list of NGC 3201 giants. We thank
Chris Smith for obtaining CCD observations of $\omega$ Cen with the CTIO
Schmidt telescope.  We acknowledge the help of C. Pryor, H. Richer, P.
Stetson, D. vandenBerg, and G. Wallerstein for discussions on some of the
points raised in this work.  NBS gratefully acknowledges the hospitality of
the Dominion Astrophysical Observatory where much of this paper was written.
RPK acknowledges the support from the NSF, under grant AST 92-17970.


\clearpage

\appendix
\section{Appendix}

Just as this paper was being completed and prepared for publication, we
received a preprint from Norris, Freeman, \& Mighell on the metallicity
distribution of $\omega$ Cen based on 518 giants on the upper RGB ($M_V \ga
-1$) using spectra of the \ion{Ca}{2} H+K and triplet lines. For 138 stars in
common between our BG sample and their sample, we find excellent agreement
between the measured \ion{Ca}{2} infrared triplet line strengths for all the
stars except ROA 320, 336, 383, and 406 where the W(Ca) values differed by
more than 0.7\AA. For the remaining 133 stars, a simple linear regression
between the two W(Ca) data sets gave a dispersion of 0.15\AA. The Norris et
al. study chose to calibrate the data with respect to the [Ca/H] values in NDC
(see the left-hand panel of Figure~\ref{x4}). A quadratic correlation
between the values of [Fe/H]$_{NDC}$ in Table~\ref{y3} as a function of the
[Ca/H] values given by Norris et al. has a dispersion of only 0.055 dex in
[Fe/H].  Evidently both W(Ca) data sets yield consistent metallicities.  The
Norris, et al. study finds a secondary hump in the metal rich part of the
metallicity distribution which we do not recover in our data. They
interpret this second hump as a second generation of stars forming from the
enriched ejecta of the primary metal-poor population. While we do not find a
secondary hump in our SG sample, we do concur that a simple model for chemical
enrichment does not predict the number of metal-rich stars observed.

\clearpage

\section*{References}
\begin{verse}
Alcaino, G. \& Liller W.  1984, AJ, 89, 1712\\
Armandroff, T.E. 1989, \aj, 97, 375\\
Armandroff, T.E, \& Da Costa, G.S. 1991, AJ, 101, 1329\\
Armandroff, T.E., Da Costa, G.S., \& Zinn, R. 1992, AJ, 104, 164 \\
Bergbusch, P.A., \& VandenBerg, D.A. 1992, \apjs, 81, 163\\
Bessell, M.S., \& Norris, J.E. 1976, \apj, 208, 369  (erratum 210, 618)\\
Binney, J. \& Tremaine, S. 1987, in Galactic Dynamics (Princeton: Princeton
 Univ. Press), 514\\
Bond, H.E. 1981, \apj, 248, 606\\
Brewer, J. P., Fahlman, G. G.,Richer, H. B., Searle, L., \& Thompson, I.
 1993, \aj, 105, 2158\\
Brown, J.H., Burkert, A., Truran, J.W. 1991, \apj, 376, 115\\
Brown, J.H., Burkert, A., Truran, J.W. 1995, \apj, 440, 666\\
Brown, J.A. \& Wallerstein, G.W. 1993, \aj, 106, 133\\
Butler, D.,  Dickens, R.J., \& Epps, E.  1978, \apj, 225, 148\\
Cannon, R.D. \& Stobie R.S.  1973, MN, 162, 207\\
Cannon, R.D.  1974, MN, 167, 551\\
Carbon, D. F., Langer, G. E., Butler, D., Kraft, R. P., Suntzeff, N. B. 
 Kemper, E., Trefzger, C. F.,  Romanishin, W., 1982, \apjs, 49
 207 \\
Cayrel, R. 1986, \aap, 168. 18\\
Chun, M.S., \& Freeman, K.C. 1978, \aj, 83, 376\\
Cohen, J.G. 1981, \apj, 247, 869\\
C\^ote, P., Welch, D. L., Fischer, P., Da Costa, G. S., Tamblyn, P.
 Seitzer, P., \& Irwin, M. J. 1994, \apjs, 90, 83\\
Cudworth, K.M., \& Rees, R. 1990, \aj, 99, 1491\\
Cudworth, K.M. 1994, private communication\\
Da Costa, G.S., Armandroff, T.E., \& Norris, J.E. 1992, \aj, 104, 154\\
Da Costa, G.S. \& Villumsen, J.E. 1981, in Astrophysical Parameters for
 Globular Clusters, eds A.G.D. Philip \& D.S. Hayes, (Schenectady; Davis)
 527\\
Dickens, R.J., Feast, M.W., \&  Lloyd Evans T. 1972, \mnras, 159, 337 \\ 
Dickens, R.J., \& Saunders, J. 1965, Roy. Obs. Bull., No. 101\\
Eggen, O.J. 1972, \apj, 172, 639\\
Erdelyi-Mendes, M. \& Barbuy, B. 1991, \aap, 241, 176\\
Fall, S.M. \& Rees, M.J. 1985, \apj 298, 18\\
Freeman, K.C. 1985, in IAU Symp. 113, p. 33\\
Francois, P, Spite, M., \& Spite, F.  1987, \aap, 191, 267\\
Fusi Pecci, F., Ferraro, F. R., Crocker, D. A., Rood, R. T., \& Buonanno, R.
 1990, \aap, 238, 95\\
Gingold, R.A. 1974, \apj, 193, 177\\
Gratton, R.G. 1982, \aap, 115, 336\\
Gunn, J.E. 1980, in Globular Clusters, eds D. Hanes \& B. Madore
 (Cambridge: Cambridge) 301\\
Hartwick, F.D.A. 1976, \apj, 209, 418\\
Hartwick, F.D.A. 1983, Mem. S.A. It., 54, 51\\
Hawarden, T.G. \&  Epps Bingham, E.A. 1987, SAAO Circ, 11, 83\\
Hesser, J.E. Hartwick, F.D.A., \& McClure, R.D. 1977, \apjs, 33, 471 \\
Kraft, R.P. 1994, \pasp, 106, 553\\
Langer, G.E., Suntzeff, N.B., \& Kraft, R.P. 1992, \pasp, 104, 523\\
Lee, S.-W. 1977a, \aaps, 27, 367 \\ 
Lee, S.-W. 1977b, \aaps, 27, 381  \\ 
Lee, S.-W. 1977c, \aaps, 28, 409 \\ 
Lloyd Evans, T. 1974, \mnras, 167, 393\\
Lloyd Evans, T. 1983a, SAAO Circ. 7, 86\\
Lloyd Evans, T. 1983b, \mnras, 204, 975\\
Mallia, E.A.,  \& Pagel, B.E.J. 1981, \mnras, 194, 421\\
Martin, W.L. 1981, SAAO Circ, 6, 28\\
Matteucci, F. 1983, Mem. S.A. It., 54, 289\\
Mayor, M., Imbert, M., Andersen, J., Ardeberg, A., Baranne, A., Benz, W.,
 Ischi, E., Lindgren, H., Martin, N., Maurice, E., Nordstr\"{o}m, B., \&
 Pr\'{e}vot, L. 1983, \aaps, 54, 495\\
Meylan, G. \& Mayor, M. 1986, \aap, 166, 122\\
Morgan, S., \& Lake, G. 1989, \apj, 339, 171\\
Nemec, J.M., \& Harris, H.H. 1987, \apj, 316, 172\\
Norris, J. 1980, in Globular Clusters, eds. D. Hanes \& B. Madore (Cambridge:
 Cambridge), 113\\
Norris, J. \& Smith, G.H.  1983, \apj, 272, 635\\
Norris, J. \& Bessell, M.S. 1975, \apj, 201, L75 (erratum in 210, 618)\\
Norris, J. \& Bessell, M.S. 1977, \apj, 211, L91\\
Norris, J.E., \& Da Costa, G.S. 1995, \apj, 447, 680\\
Olszewski, E. W., Schommer, R. A., Suntzeff, N. B., \& Harris, H. C. 1991,
 \aj, 101, 515\\
Pagel, B.E.J. 1992, in The Stellar Populations of Galaxies, IAU Symp. 149,
 eds. B. Barbuy \& A. Renzini (Dordrecht: Kluwer) 133\\
Paltoglou, G. \& Norris, J.E. 1989, \apj, 336, 185\\
Persson, S. E., Cohen, J. G.,  Matthews, K., Frogel, J. A., \& 
 Aaronson, M.  1980, \apj, 235,  452\\
Peterson, R. C., Rees, R. F.,  Cudworth, K. M.   1995 \apj, 443, 124\\
Press, W.H., Teukolsky, S.A., Vetteriling, W.T., \& Flannery, B.P. 1992,
 Numerical Recipes, (Cambridge: Cambridge, MA), p. 617\\
Pryor, C. \& Meylan, G. 1993, in Structure and Dynamics of Globular Clusters,
 ASP Conf. Series Vol 50, eds. S.G. Djorgovski \& G. Meylan (San Francisco:
 ASP) 357\\
Richer, H.B., Fahlman, G.G., Buonanno, R., Fusi Pecci, F., Searle, L., \&
 Thompson, I.B. 1991, \apj, 381, 147\\
Sandage. A.R., \& Cacciari, C. 1990, \apj, 350, 645\\
Searle, L., \& Sargent, W.L.W. 1972, \apj, 173, 25\\
Smith, G.E. 1984, \aj, 89, 801\\
Stetson, P.B. 1994, \pasp, 106, 250\\
Suntzeff, N.B. 1981, \apjs, 47, 1\\
Suntzeff, N.B. 1993, in the Globular Cluster-Galaxy Connection,
 ASP Conf. Series, Vol 48, eds G.H. Smith \& J.P. Brodie (San Francisco: ASP) 
 167\\
Suntzeff, N.B., Kraft, R.P., \& Kinman, T.D.  1994, \apjs, 93, 271\\
Suntzeff, N. B., Schommer, R. A., Olszewski, E. W., \& Walker, A. R. 1992,
 \aj, 104, 1743\\
Suntzeff, N.B., Mateo, M.,  Terndrup, D.M., Olszeswki, E.W.,  Geisler, D., 
 \& Weller, W.  1993, \apj, 418, 208 (S93)\\
Trager, S.C., King, I.R., \& Djorgovski, S. 1995, \aj, 109, 218\\
Vandenberg, D.A. 1991, \apj, 391, 685\\
Vietri, M. \& Pesce, E. 1995, \apj, 442, 618\\
Webbink, R.F. 1985, in IAU Symp. 113, Dynamics of Star Clusters,
 ed. J. Goodman \& P. Hut (Dordrecht: Reidel) 541 \\
Wildey, R. 1961, \apj, 133, 430\\
Woolley, R. v.d.R.  1966. Royal Obs. Ann., No. 2\\
Woolley, R.v.d.R., Alexander, J.B., Mather, L., \& Epps, E. 
 1961, Roy. Obs.  Bull. No. 43\\
Zinn, R. 1978, \apj, 225, 790\\
Zinn, R., \& West, M.J. 1984, ApJS, 55, 45\\
\end{verse}

\clearpage
\section*{Figure Captions}

\figcaption[]{A color-magnitude diagram showing the two sample regions
in $\omega$ Cen observed in this study. The $BV$ photometry is from Woolley
(1966). Only stars with small proper motions are plotted. \label{x1}}

\figcaption[]{A color-magnitude diagram of $\omega$ Cen showing the stars
observed in this study. The filled circles represent radial velocity members
and the exes non-members. The $BV$ photometry is from Woolley
(1966). \label{x6}}

\figcaption[]{The error in a single observation plotted as a function of
the logarithm of the mean intensity in the co-added spectrum. The upper plot
presents the data for the W(Ca) errors in units of (\AA), and the lower
plots presents the velocity errors in units of km s$^{-1}$. The curves
represent the median points in the distributions. \label{x8}}

\figcaption[]{The \protect{\ion{Ca}{2}} infrared triplet equivalent 
width \wca\ in units of (\AA) plotted as a function of the $V$ magnitude with
respect to the cluster horizontal branch for giants in the calibrating
clusters. \label{x2}}

\figcaption[]{The \protect{\ion{Ca}{2}} infrared triplet equivalent width \wca\ in units
of (\AA) for the bright and faint star sample plotted as a function of
\vminusvhb. \label{x7}}

\figcaption[]{The cluster [Fe/H] from Zinn \& West (1984) plotted as a
function of the averaged ``reduced'' equivalent widths in units of (\AA) for
the calibrating clusters (1984). The closed circles are data from this work
(NGC 3201) or S93, and the open circles are from Armandroff \etal\ (1992) and
Armandroff \& Da Costa (1991). The solid lines are the best fits providing a
continuous relationship which were used to calculate ${\rm [Fe/H]}_{ZW}$ for
the $\omega$ Cen giants. \label{x3}}

\figcaption[]{A comparison of the reduced equivalent width $W'$ 
in units of (\AA) with the calcium and iron abundances in individual stars in
$\omega$ Cen. The detailed abundances were taken from Norris \& Da Costa
(1995). The dotted line is the relationship given in the text to calculate
${\rm [Fe/H]}_{NDC}$. The open circles are some other globular cluster stars
observed by Norris \& Da Costa (1995) for which we also have $W'$
values. \label{x4}}

\figcaption[]{Spectra of the \protect{\ion{Ca}{2}} infrared triplet region for the SG 
sample $\omega$ Cen giants showing the range in line strength. From top to
bottom the stars are ROA 1317 ([Fe/H]$_{NDC}=-1.81$), ROA 5612 (--1.30), and
ROA5017 (--1.04).  All the spectra have been normalized to a continuum value
of 1 and shifted by increments of 0.5. \label{x5}}

\figcaption[]{Dereddened $(B-V)_0$ values plotted as a function of
[Fe/H] for the SG sample. The data are plotted on the ZW scale (left panel)
and the NDC scale (right panel). The relationship for $(B-V)_{0,g}$ as a
function of [Fe/H] from Webbink (1985) is shown as the solid line. The range
in metallicity is better fit by the NDC scale. \label{x12}}

\figcaption[]{The multiplicative correction factor to correct the observed
number of stars binned as a function of metallicity in the SG or BG sample to
an equivalent number of main sequence stars just below the turn off. The solid
line is the correction for the SG sample and the dotted line is the correction
for the BG sample. These correction factors were calculated for a mass
function exponent of $x=1$. \label{x9}}

\figcaption[]{The metallicity histograms for the cluster $\omega$ Cen.
These histograms have been corrected for evolutionary effects, and represent
the relative number of stars between 0.6 and $0.7M_{\sun}$ as a function of
metallicity. The solid-line histogram represents the SG sample and the
dotted-line histogram the BG sample. The left panel is on the Zinn \& West
(1984) metallicity scale, and the right panel is on the Norris \& Da Costa
(1995) scale. The BG sample has been scaled to the sample size of the SG
sample to allow comparison. \label{x10}}

\figcaption[]{Plot of radial gradient (in units of arcminutes) 
of metallicity in $\omega$ Cen. The left panel shows the BG sample (solid
circles) and the Seitzer bright giants (open circles). The right panel shows
the SG sample. The metallicities are plotted on the NDC scale. \label{x11}}

\figcaption[]{Histogram of the metallicity distribution for the SG sample 
compared to an example fit to a simple model of chemical evolution.  The solid
curve is the histogram of observed [Fe/H] values on the [Fe/H]$_{NDC}$
scale. The dotted line curve is the model (on an arbitrary vertical scale),
and the solid line curve is the model convolved with a gaussian of dispersion
of 0.07 dex in [Fe/H]. The dotted histogram is the histogram of the convolved
model scaled to the observed data. \label{x13}}

\clearpage

\begin{table}
\dummytable\label{y1}
\end{table}
\begin{table}
\dummytable\label{y2}
\end{table}
\begin{table}
\dummytable\label{y3}
\end{table}
\begin{table}
\dummytable\label{y4}
\end{table}
\begin{table}
\dummytable\label{y5}
\end{table}
\begin{table}
\dummytable\label{y7}
\end{table}
\begin{table}
\dummytable\label{y6}
\end{table}
%

\noindent{\sc TABLE}~\ref{y1}.  Globular Clusters Used for Calibration

\noindent{\sc TABLE}~\ref{y2}.  Photometry, EQW values, and Radial Velocities
 for Standard Clusters

\noindent{\sc TABLE}~\ref{y3}a. Photometry, EQW values, and Radial Velocities
 for$\omega$\ Cen BG Sample

\noindent{\sc TABLE}~\ref{y3}b. Photometry, EQW values, and Radial Velocities 
for $\omega$\ Cen SG Sample

\noindent{\sc TABLE}~\ref{y4}.   $\omega$ Cen Velocity Non-Members

\noindent{\sc TABLE}~\ref{y5}. Velocity Comparisons

\noindent{\sc TABLE}~\ref{y7}.  Median Values for the Error in a Single 
 Observation

\noindent{\sc TABLE}~\ref{y6}.  Metallicity Averages for $\omega$ Cen

\clearpage

\begin{figure}
\plotone{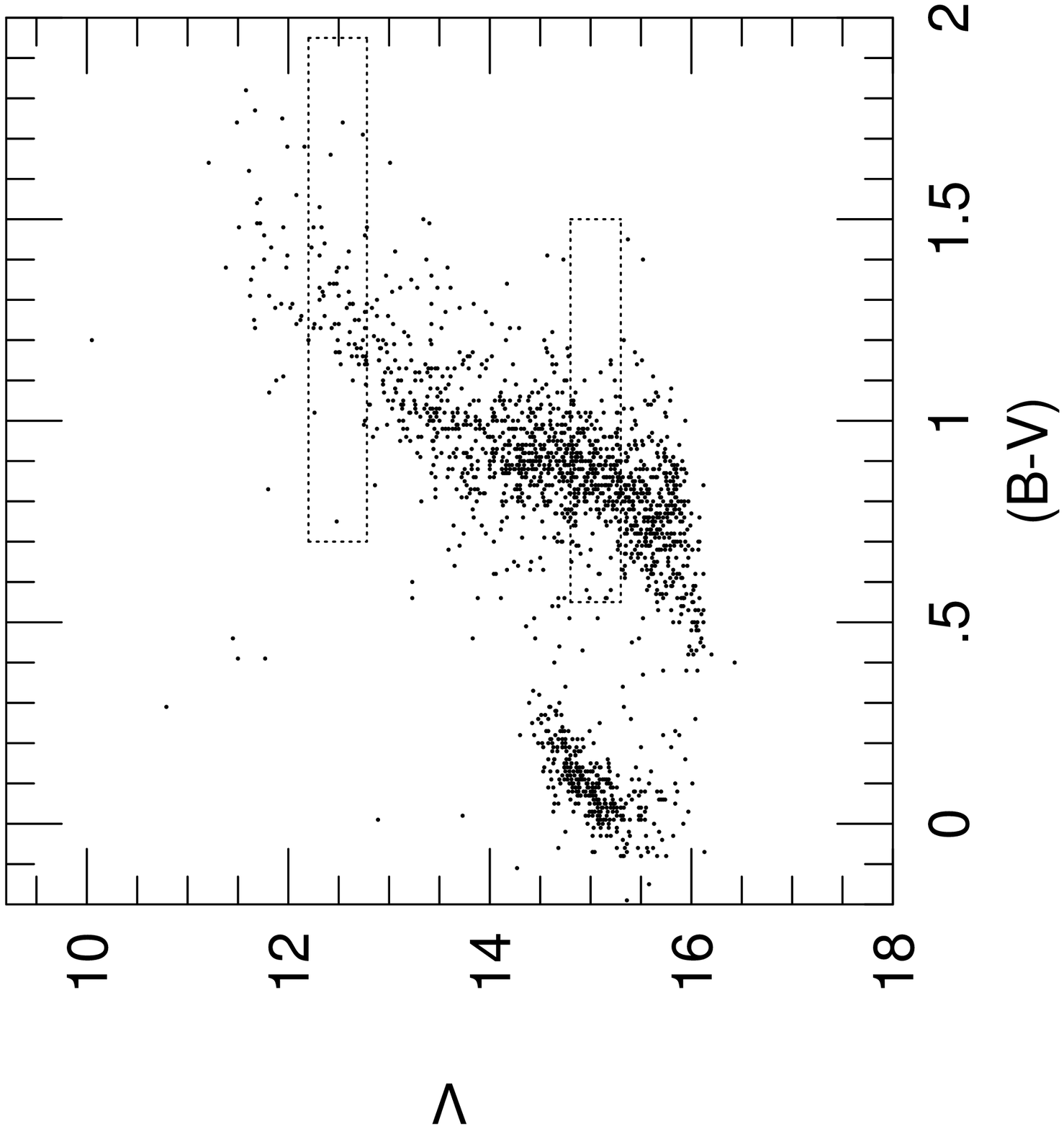}
{\center Suntzeff and Kraft. Figure~\ref{x1}}
\end{figure}

\begin{figure}
\plotone{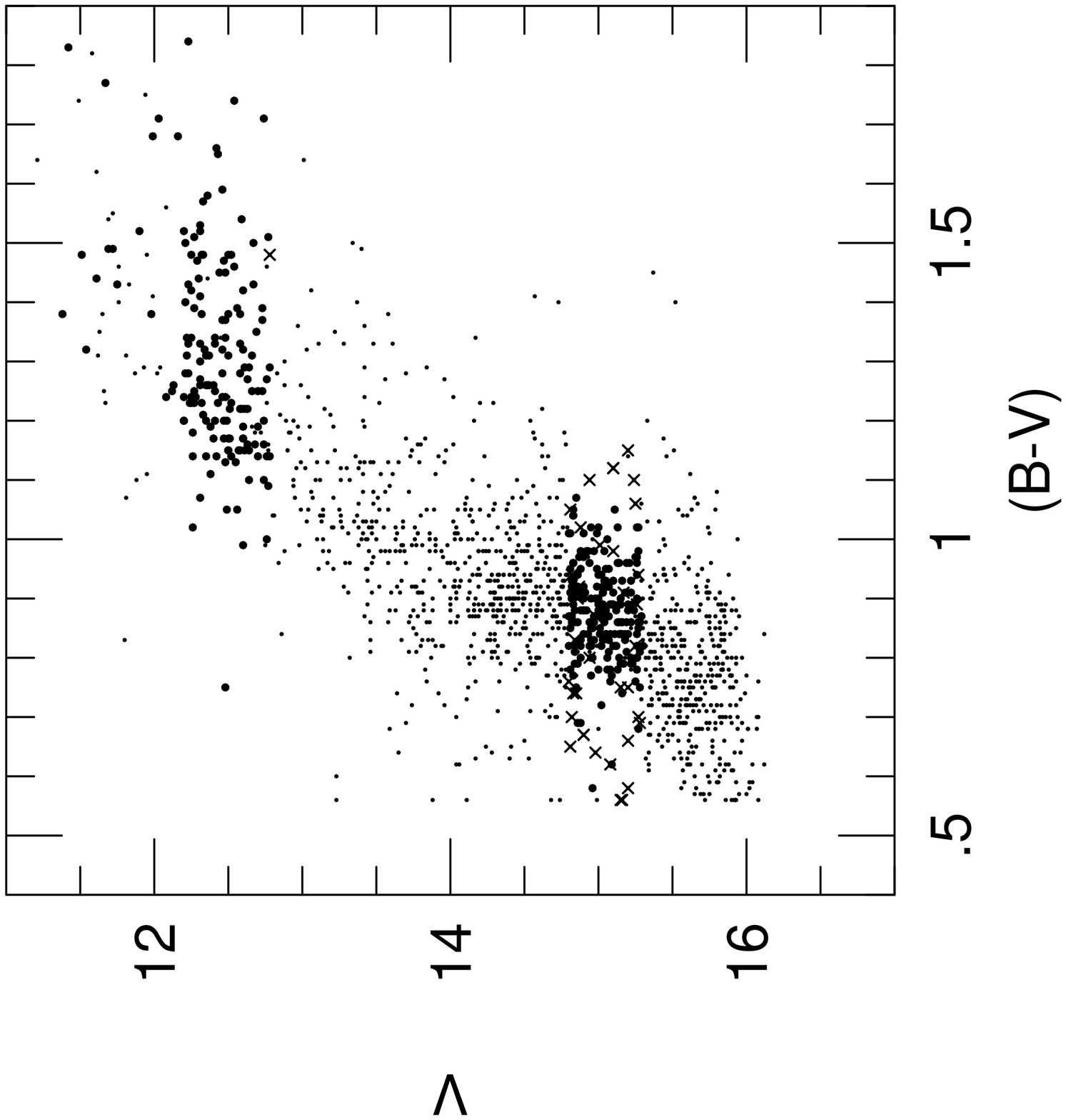}
{\center Suntzeff and Kraft. Figure~\ref{x6}}
\end{figure}

\begin{figure}
\plotone{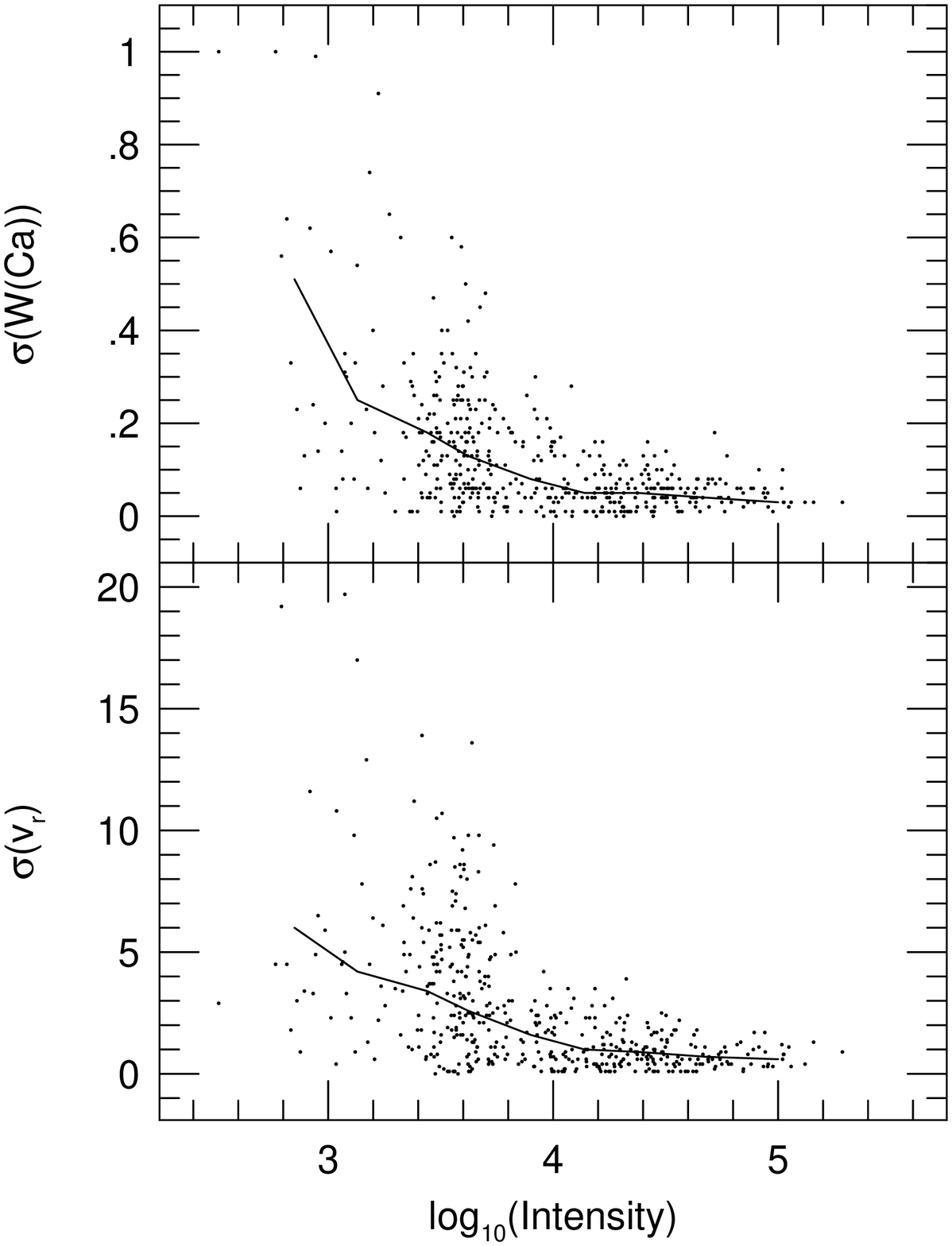}
{\center Suntzeff and Kraft. Figure~\ref{x8}}
\end{figure}

\begin{figure}
\plotone{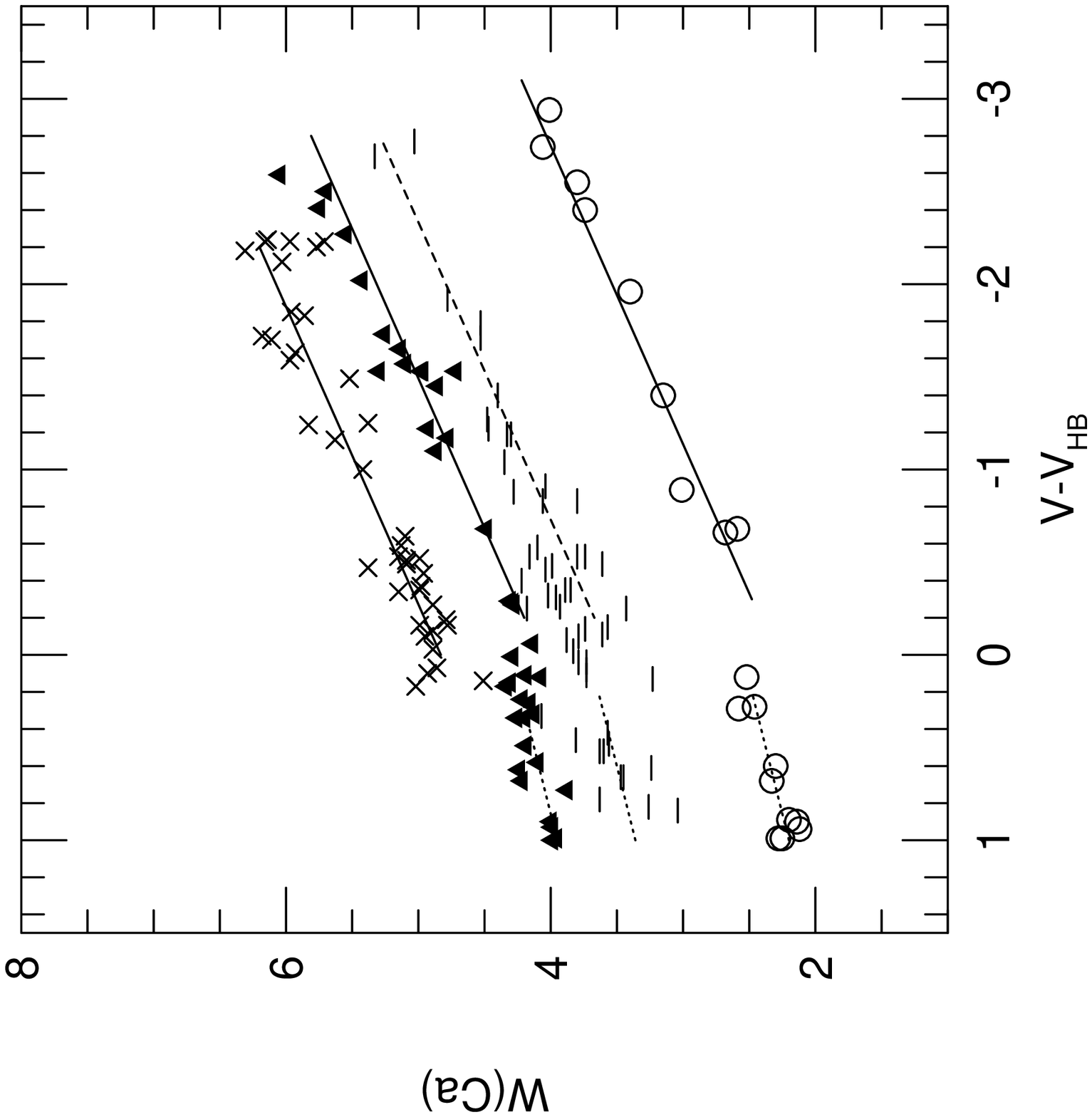}
{\center Suntzeff and Kraft. Figure~\ref{x2}}
\end{figure}

\begin{figure}
\plotone{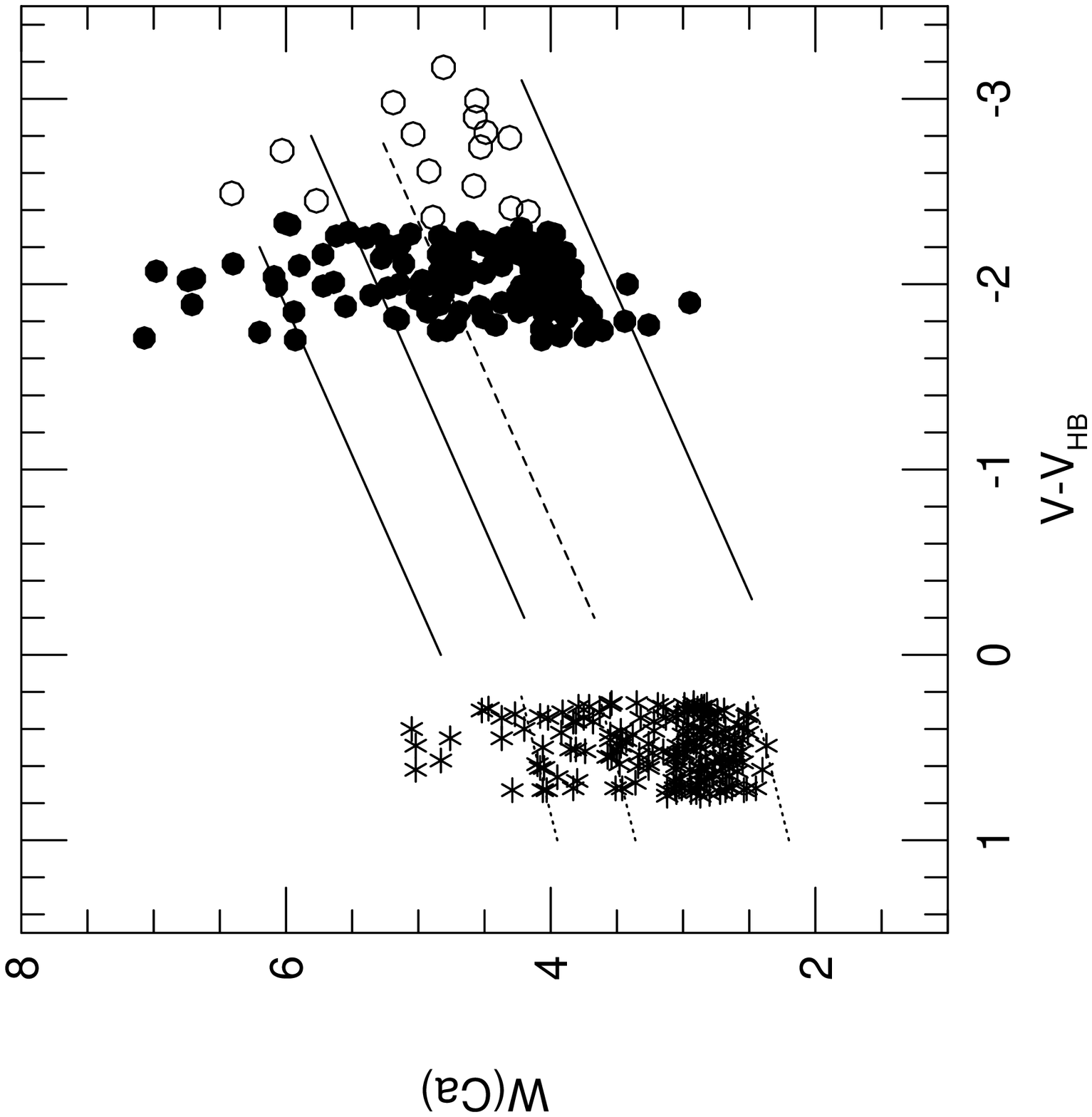}
{\center Suntzeff and Kraft. Figure~\ref{x7}}
\end{figure}

\begin{figure}
\plotone{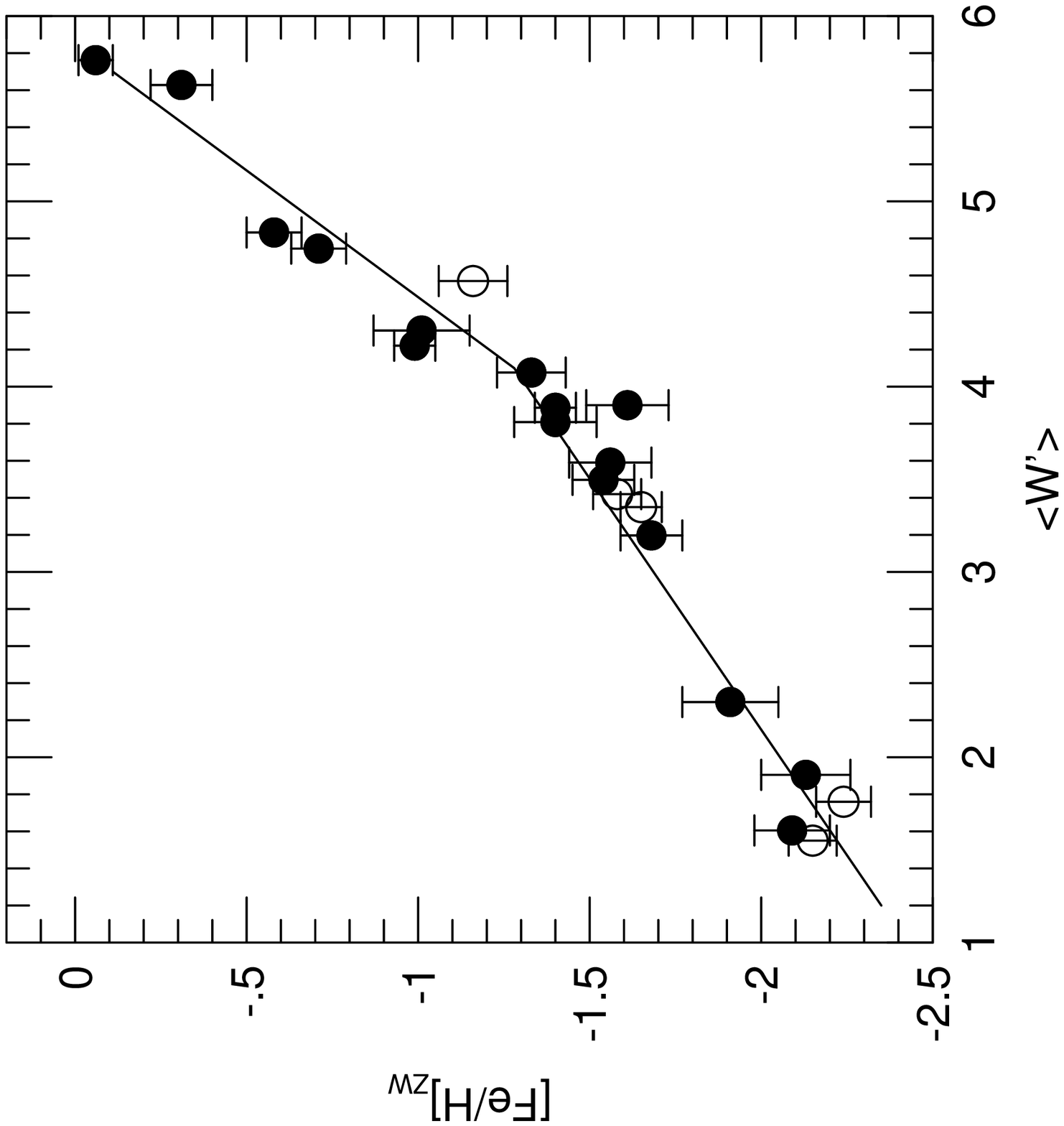}
{\center Suntzeff and Kraft. Figure~\ref{x3}}
\end{figure}

\begin{figure}
\plotone{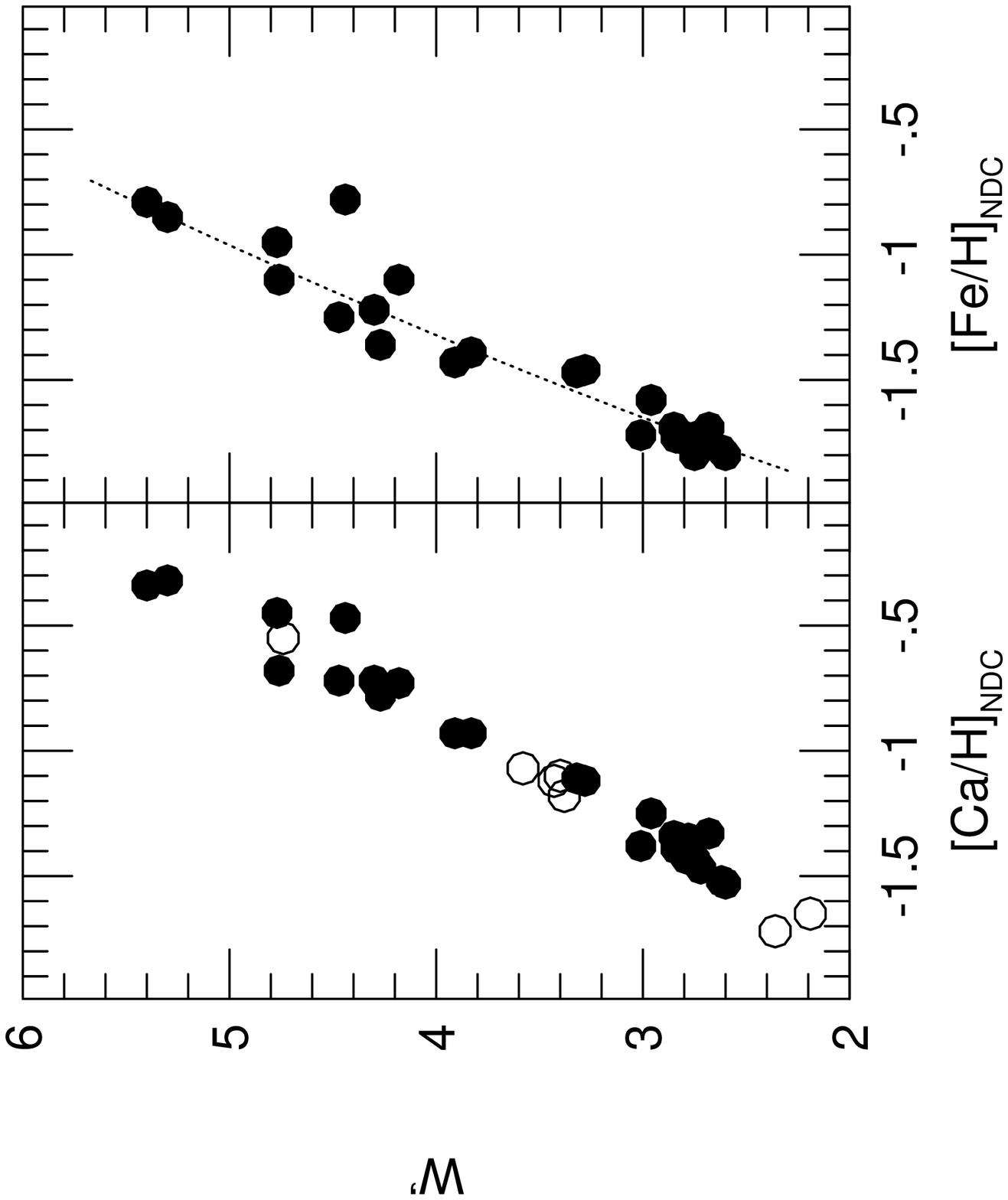}
{\center Suntzeff and Kraft. Figure~\ref{x4}}
\end{figure}

\begin{figure}
\plotone{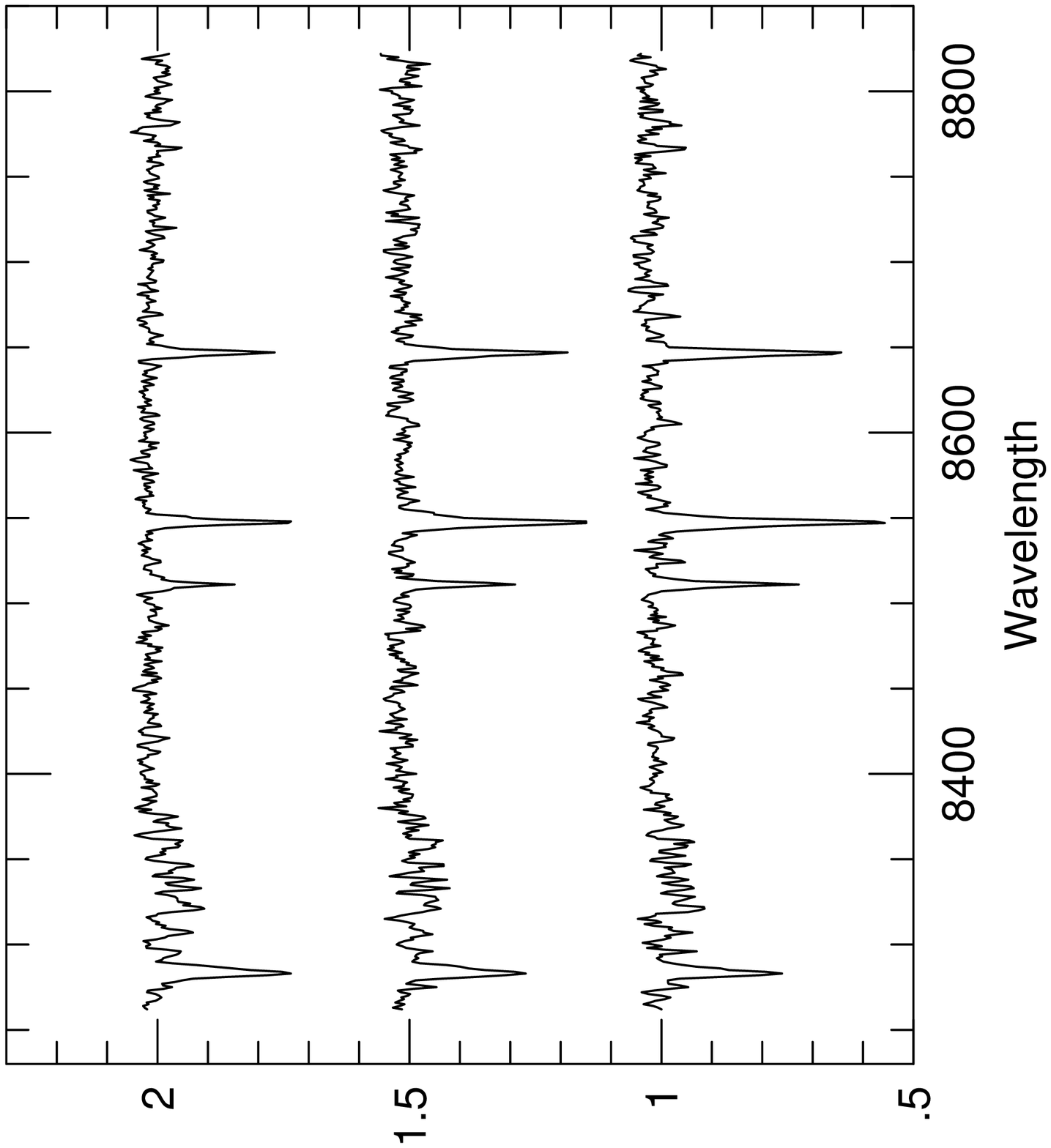}
{\center Suntzeff and Kraft. Figure~\ref{x5}}
\end{figure}

\begin{figure}
\plotone{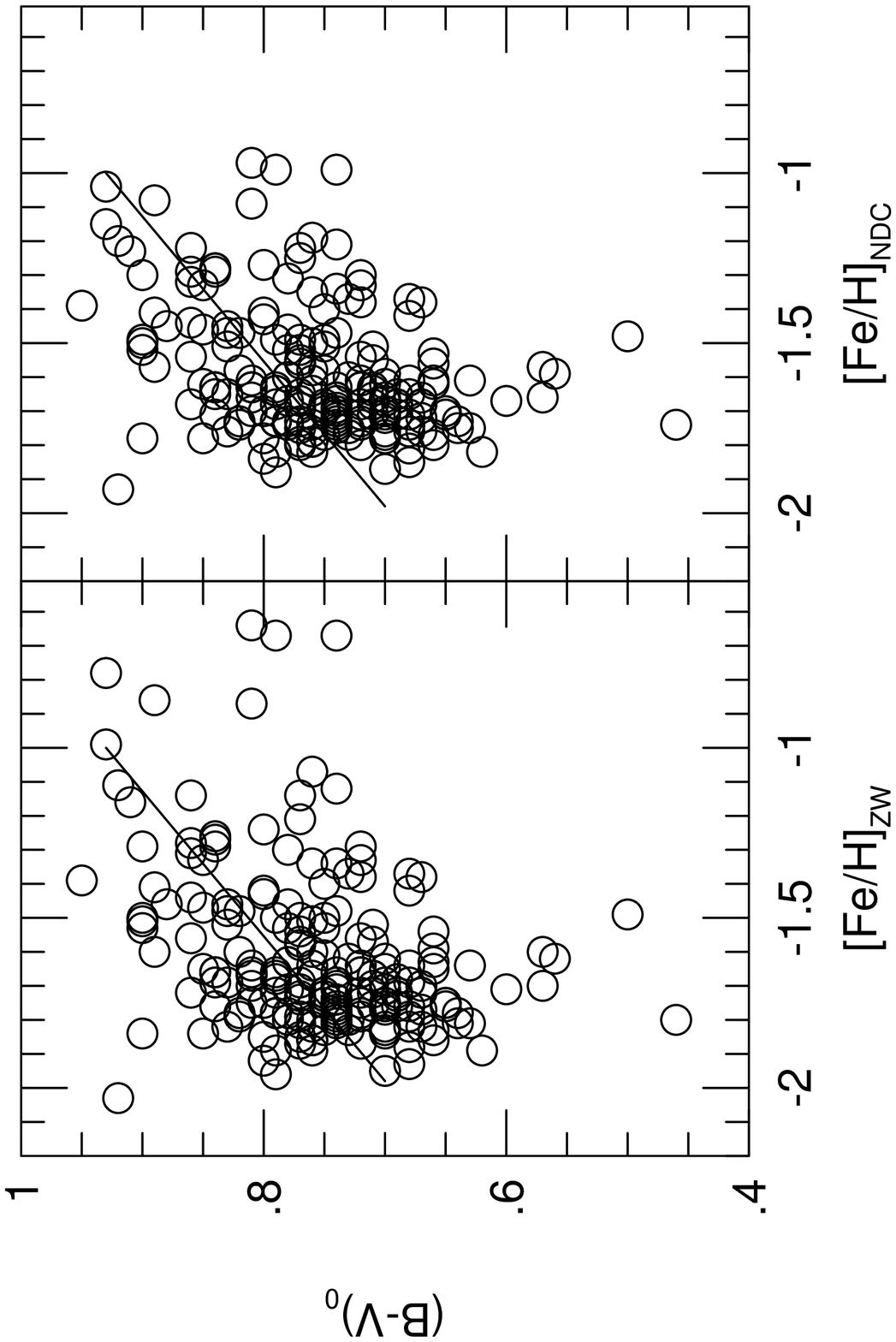}
{\center Suntzeff and Kraft. Figure~\ref{x12}}
\end{figure}

\begin{figure}
\plotone{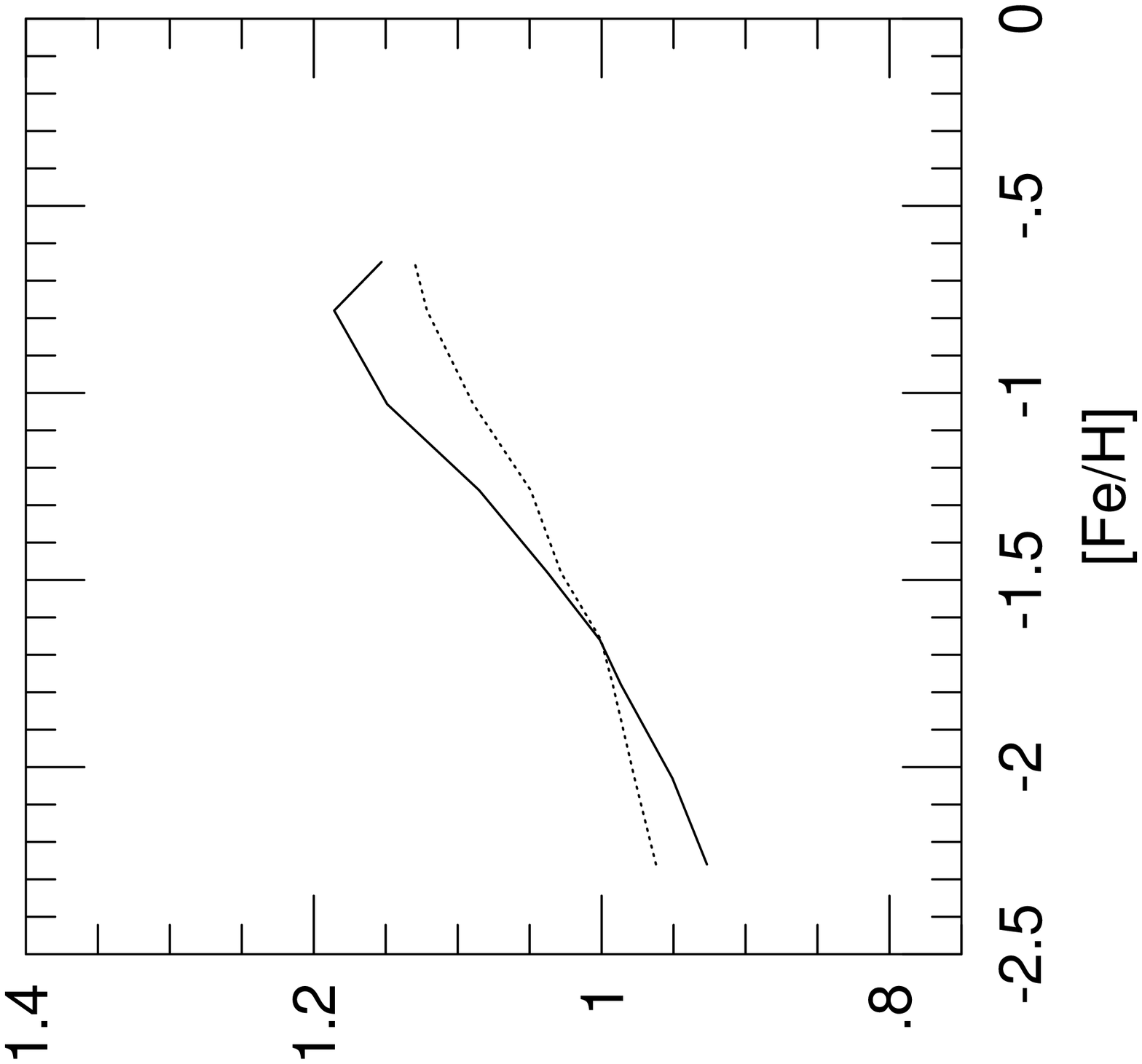}
{\center Suntzeff and Kraft. Figure~\ref{x9}}
\end{figure}

\begin{figure}
\plotone{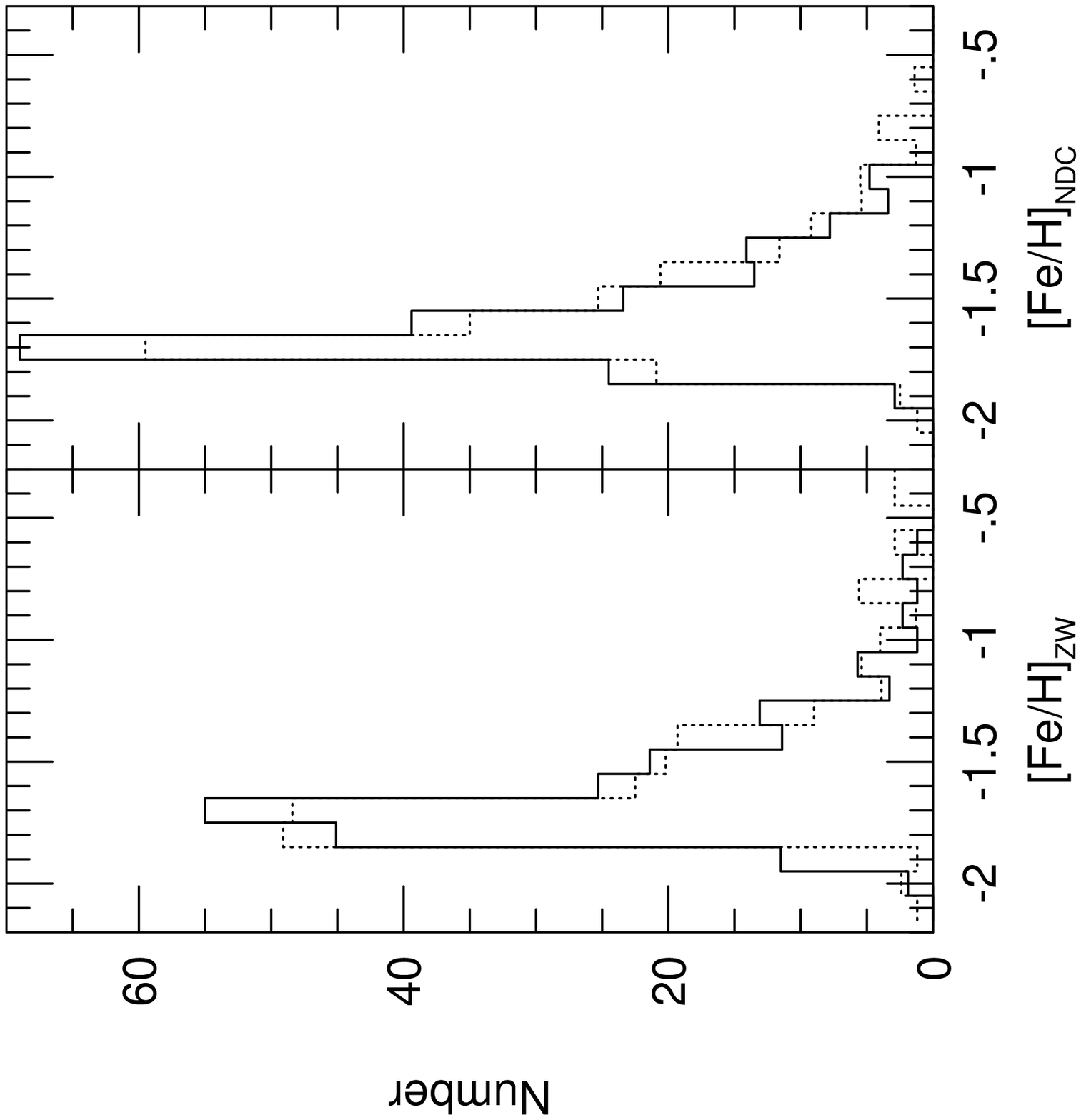}
{\center Suntzeff and Kraft. Figure~\ref{x10}}
\end{figure}

\begin{figure}
\plotone{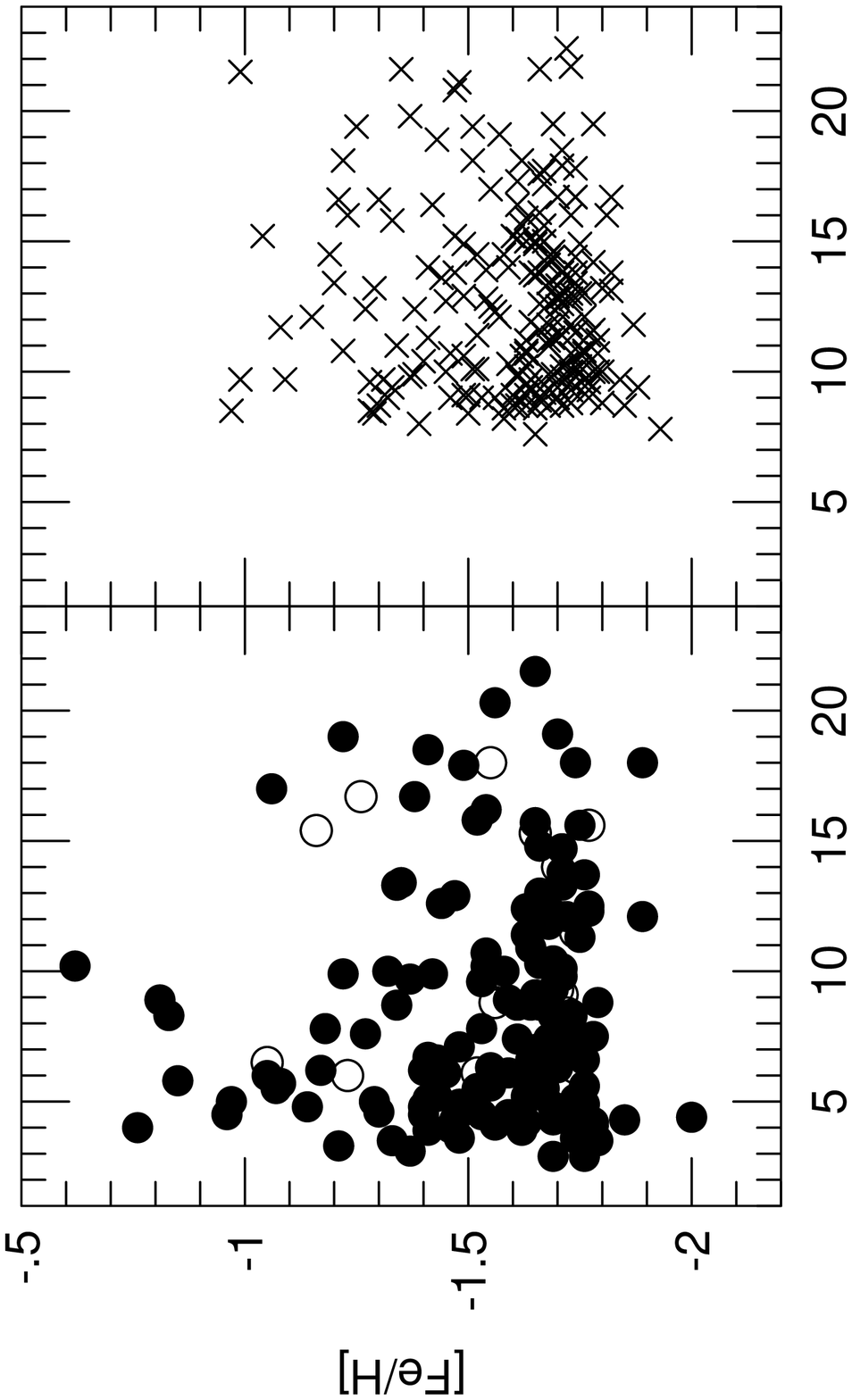}
{\center Suntzeff and Kraft. Figure~\ref{x11}}
\end{figure}

\begin{figure}
\plotone{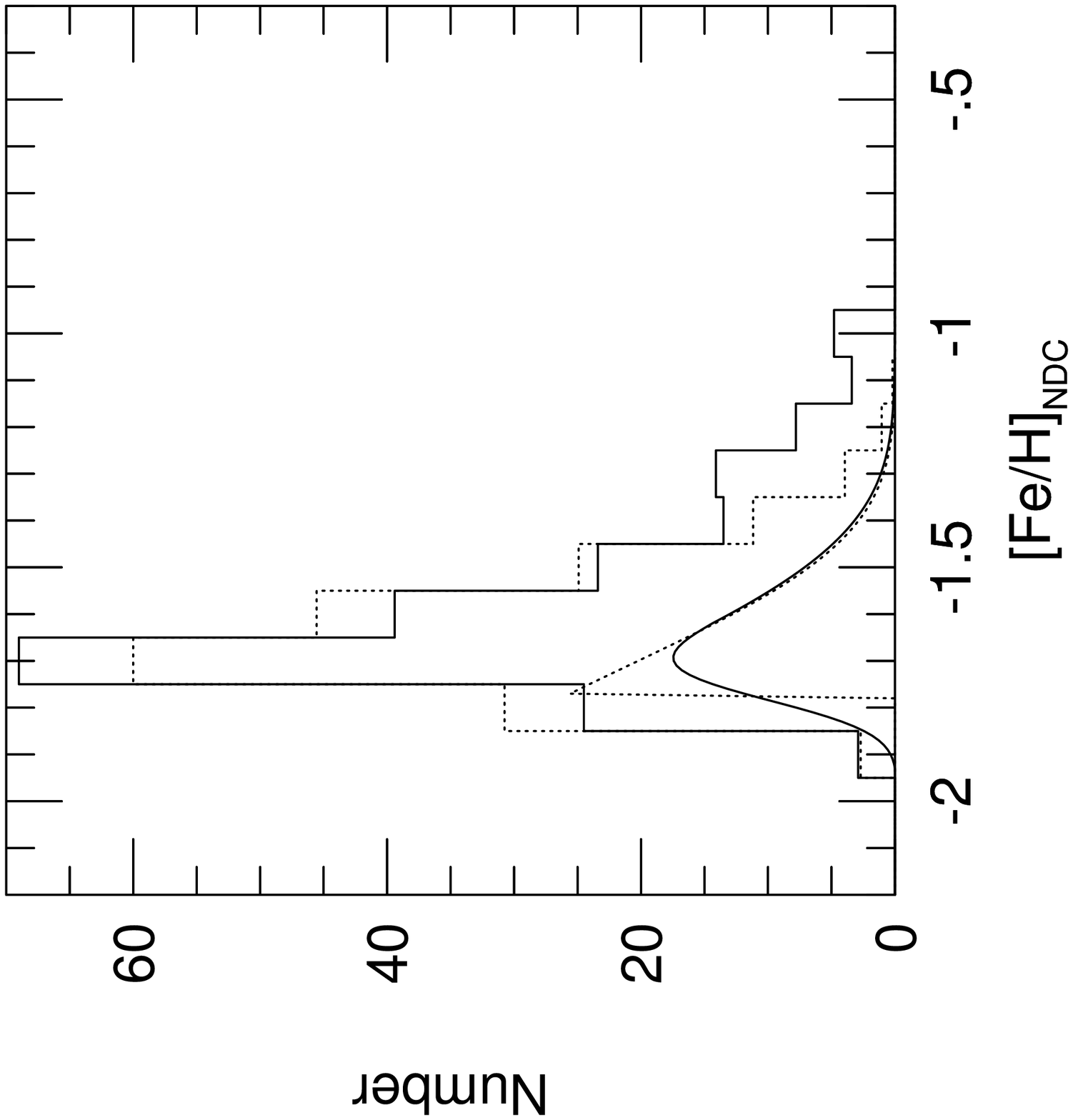}
{\center Suntzeff and Kraft. Figure~\ref{x13}}
\end{figure}

\end{document}